\documentclass[lettersize,journal]{IEEEtran}
\usepackage{amsmath,amsfonts}
\usepackage{booktabs}
\usepackage{multirow}
\usepackage{makecell}
\usepackage{algorithmic}
\usepackage{algorithm}
\usepackage{array}
\usepackage[caption=false,font=normalsize,labelfont=sf,textfont=sf]{subfig}
\usepackage{textcomp}
\usepackage{stfloats}
\usepackage{url}
\usepackage{verbatim}
\usepackage{graphicx}
\usepackage{cite}
\begin{document}

\title{Deep Neural Network for Blind Visual Quality Assessment of 4K Content}

\author{Wei Lu, Wei Sun, Xiongkuo Min,~\IEEEmembership{Member,~IEEE,} Wenhan Zhu, Quan Zhou, Jun He, Qiyuan Wang, \\Zicheng Zhang, Tao Wang, Guangtao Zhai,~\IEEEmembership{Senior Member,~IEEE}
\thanks{Wei Lu, Wei Sun, Xiongkuo Min, Wenhan Zhu, Zicheng Zhang, Tao Wang, and Guangtao Zhai are with the Institute of Image Communication and Network Engineering, Shanghai Jiao Tong University, Shanghai 200240, China (e-mail:SJTU-Luwei@sjtu.edu.cn, sunguwei@sjtu.edu.cn, minxiongkuo@sjtu.edu.cn, zhuwenhan823@sjtu.edu.cn, zzc1998@sjtu.edu.cn, f1603011.wangtao@sjtu.edu.cn, zhaiguangtao@sjtu.edu.cn). Quan Zhou, Jun he, and Qiyuan Wang are with the Department of Video Cloud Technology, Bilibili Inc, Shanghai 200433, China (email:zhouquan@bilibili.com, hejun@bilibili.com, wangqiyuan@bilibili.com)}
\thanks{(Corresponding author: Guangtao Zhai.)}
}



\maketitle

\begin{abstract}
The 4K content can deliver a more immersive visual experience to consumers due to the huge improvement of spatial resolution. However, existing blind image quality assessment (BIQA) methods are not suitable for the original and upscaled 4K contents due to the expanded resolution and specific distortions. In this paper, we propose a deep learning-based BIQA model for 4K content, which on one hand can recognize true and pseudo 4K content and on the other hand can evaluate their perceptual visual quality. Considering the characteristic that high spatial resolution can represent more abundant high-frequency information, we first propose a Grey-level Co-occurrence Matrix (GLCM) based texture complexity measure to select three representative image patches from a 4K image, which can reduce the computational complexity and is proven to be very effective for the overall quality prediction through experiments. Then we extract different kinds of visual features from the intermediate layers of the convolutional neural network (CNN) and integrate them into the quality-aware feature representation. Finally, two multilayer perception (MLP) networks are utilized to map the quality-aware features into the class probability and the quality score for each patch respectively. The overall quality index is obtained through the average pooling of patch results. The proposed model is trained through the multi-task learning manner and we introduce an uncertainty principle to balance the losses of the classification and regression tasks. The experimental results show that the proposed model outperforms all compared BIQA metrics on four 4K content quality assessment databases.
\end{abstract}


\section{Introduction}
\IEEEPARstart{D}{ue}  to the great advancements in hardware technology of multimedia and the increasing demands for better quality of experience (QoE) of users, recent years have witnessed a growing popularity of Ultra High-Definition (UHD) content with a resolution of 3840 × 2160 pixels, also known as 4K content. As the next stage of High-Definition (HD) technology, UHD technology can deliver a more realistic and immersive visual experience to viewers \cite{cheon2017subjective}. As a result, 4K content has become a hot topic, followed by a great deal of research. 

Nowadays, most broadcasters and online video websites could provide video content at 4K resolution. Besides, the electronic devices which support 4K display are becoming more and more popular. However, with the repaid flourishing of 4K content, some problems have arisen. First, in order to take advantage of the 4K boom, many video content providers/creators utilize the upsampling or super-resolution (SR) algorithms \cite{zha2021triply,zha2019rank,zha2020image1,zha2020image2} to obtain high-resolution (HR) content interpolated from low-resolution (LR) content, which may suffer degradations in quality and can not provide the same immersive experience as true 4K content. What's more, some low-quality pseudo 4K videos bring a high cost to content service providers but deliver very poor QoE to viewers. Hence, it is important for content service providers to recognize the true and pseudo 4K content. Second, the increase in video spatial resolution inevitably leads to larger data volumes, therefore, video service providers are challenged to transmit 4K videos through limited bandwidth networks \cite{9162073}. A recent approach \cite{mackin2018study} adopted by video service providers is that the source 4K video is downsampled to meet a given bit rate requirement before encoding and upsampled after decoding, which results in a perceptually less degraded video. Therefore, it is necessary to understand how different degrees of spatial resolution downsampling affect the perceptual quality of videos. So, there is an urgent need to develop the effective quality assessment tool for HR content. 

Over the past two decades, plenty of image quality assessment (IQA) methods have been proposed. According to whether to access the reference information, IQA methods can be classified into three categories, Full Reference IQA (FR IQA), Reduced Reference IQA (RR IQA), and No Reference IQA (NR IQA) as well as known as blind IQA (BIQA) \cite{8576582}. FR IQA and RR IQA require the full and partial reference information, respectively, while BIQA does not need any reference information. The FR IQA models have been developed maturely due to the availability of the reference images. The well-known FR IQA methods such as PSNR, SSIM \cite{wang2004image}, LPIPS \cite{zhang2018unreasonable} have been successfully applied to image compression and enhancement systems. However, the reference image is usually not available in the practical applications, so, BIQA is more valuable for practical applications and attracts lots of attention in recent years \cite{zhai2021perceptual}. BIQA methods can be categorized according to distortion types, namely distortion-specific and general-purpose algorithms \cite{sun2020dynamic}. General-purpose metrics utilize the general quality features which are supposed to be able to describe various distortions, while distortion-specific metrics extract the quality features suited for the quality degradations caused by a specific distortion process. 

\begin{figure*}[!t]
\centering
\subfloat[]{\includegraphics[width=3.3in]{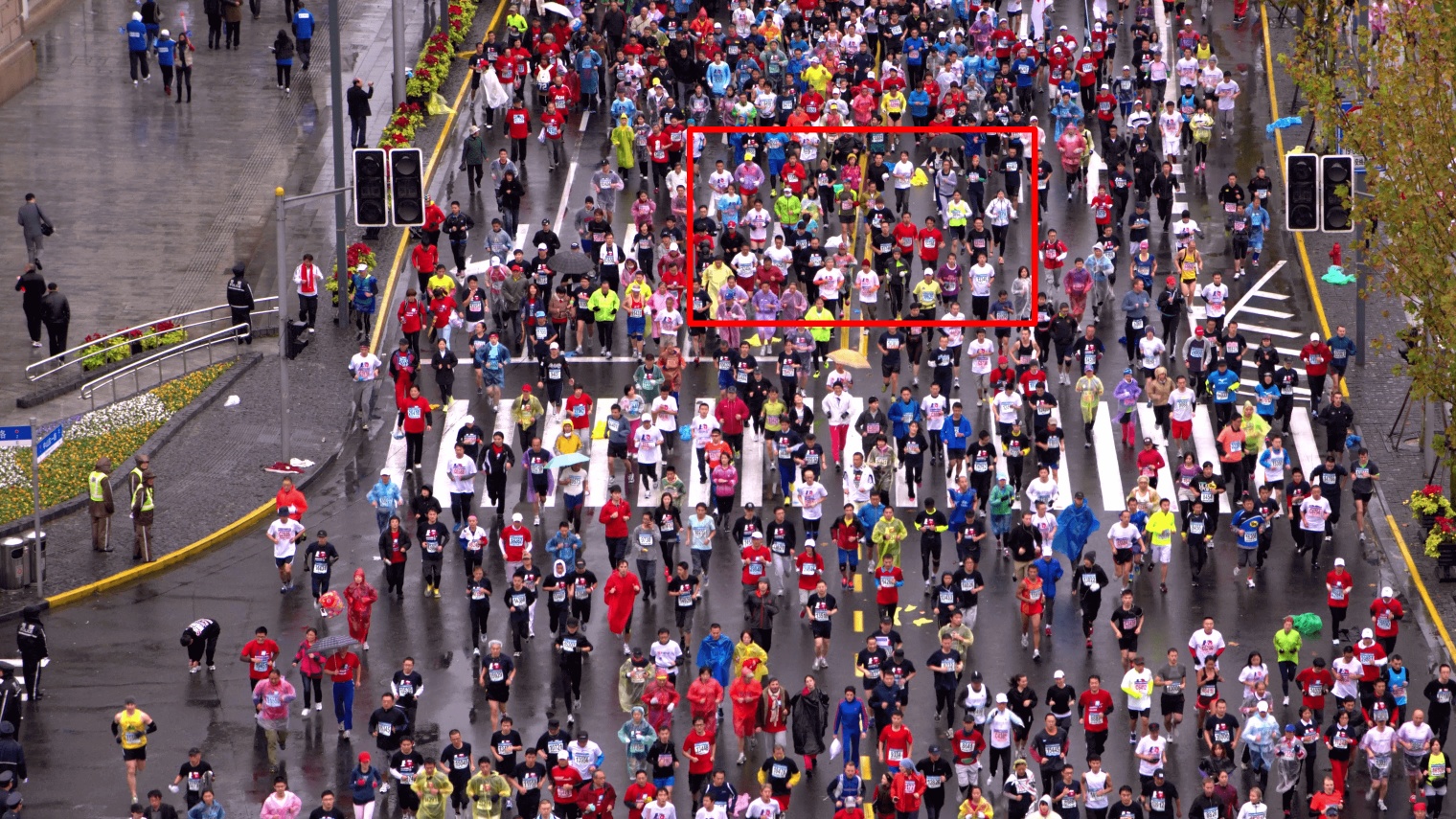}%
}
\hfil
\subfloat[]{\includegraphics[width=3.3in]{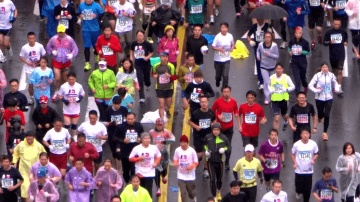}%
}
\hfil
\subfloat[]{\includegraphics[width=3.3in]{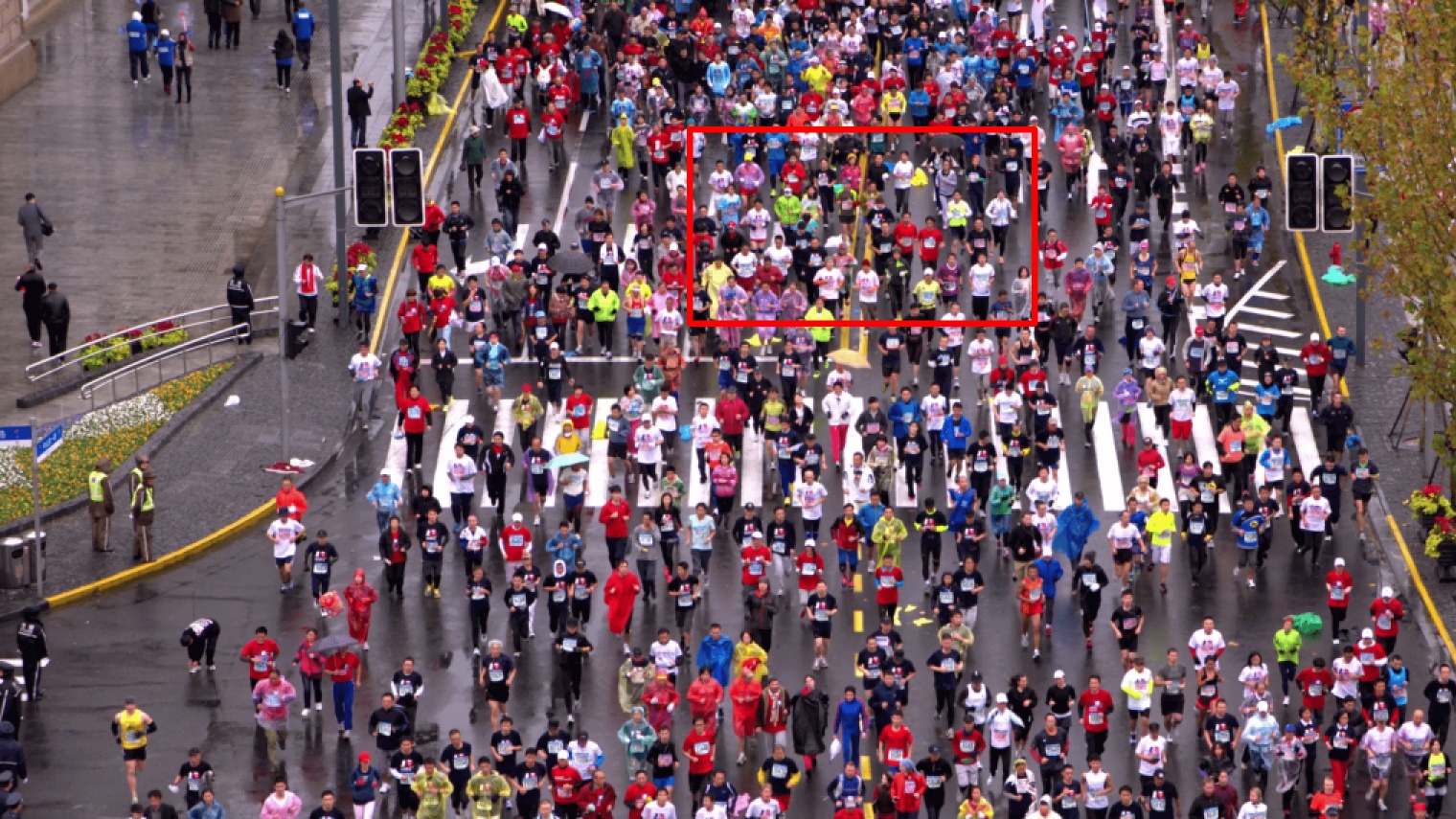}%
}
\hfil
\subfloat[]{\includegraphics[width=3.3in]{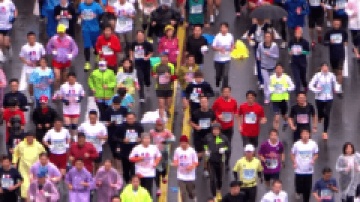}%
}
\caption{An example of quality degradations in pseudo 4K image. (a) is the true 4K image and (c) is the pseudo 4K image upscaled from 720p, (b) and (d) are the images cropped from (a) and (c) respectively.}
\label{fig1}
\end{figure*}

Compared to BIQA methods aimed at LR images, the 4K IQA research faces two main challenges: very high resolution and specific distortion types in upscaled 4K images. For the first challenge, the high spatial resolution will inevitably lead to a significant increase in computing complexity for IQA methods, which makes these methods unable to work in practical applications. For the second challenge, the pseudo 4K images usually lack the high-frequency information, which may cause textural degenerations such as dissimilar texture, checkerboard artifacts, etc, and structural distortions such as blurring and jaggies artifacts \cite{zhou2019visual}. As shown in Fig.~\ref{fig1}, the pseudo 4K image blurs the regions with rich texture, such as the red rectangle region. The existing BIQA methods \cite{mittal2012no,xue2014blind,wu2015highly,xu2016blind,7060692} are normally proposed for LR images and most of them are developed for solving the images with synthetic distortions such as JPEG compression, Gaussian noise, etc, which causes them ineffective for HR images with upscaling artifacts. It is worthy to note that another similar topic is SR IQA, which aims to evaluate the perceptual quality of SR images and the performance of SR algorithms. The difference between our 4K IQA research and SR IQA research is that, 4K IQA focus on both the original 4K images and upscaled pseudo 4K images, while existing SR IQA studies \cite{ma2017learning,zhou2019visual,kohler2019toward,beron2020blind} only involve SR images up to 1080p, which limits their applications to UHD content. To address the challenges mentioned above, two BIQA methods \cite{shah2021real,zhu2021perceptual} specifically designed for the 4K content have been proposed, which predict the quality of the selected few local patches and obtain the overall quality of the 4K content through a pooling strategy. These patch-based methods achieve lower computational complexity. 
However, the effectiveness of the patch selection strategy and the quality-aware feature extraction method have not been explored in depth. Specifically, on
the one hand, the patch selection strategy adopted by current 4K BIQA researches, is based on some simple texture measures such as variance, which can not effectively select representative local patches to reflect the overall visual quality of the whole 4K content. On the other hand, for the quality-aware features extraction methods, they usually extract hand-crafted features or directly adopt a lightweight CNN to extract features, which are not powerful to represent the visual
quality of local patches.

In order to address the problems mentioned above, we propose a 4K BIQA method named as \textbf{b}lind \textbf{t}exture-aware \textbf{U}HD content \textbf{r}ecognition and \textbf{a}ssessment (BTURA) metric, which aims to separate the true 4K images from the pseudo 4K images and measure their visual quality at the same time. The proposed BTURA metric consists of three parts: textured patch selection module, quality-aware feature extraction module, and quality evaluation module. Since the UHD image can provide more abundant image texture and details due to the huge increase of the spatial resolution, the texture-rich image regions are more representative for its visual quality. Hence, patch selection module selects three representative patches with the highest texture complexity measured by the Grey-level Co-occurrence Matrix (GLCM) \cite{haralick1973textural} based texture complexity algorithm. The patch selection operation can help reduce the computational cost of the proposed model to a great extent and the experimental results indicate that the selected patches by the adopted strategy contain sufficient information for the accurate quality prediction of 4K content. As for the feature extraction module, we extract different kinds of visual features from the intermediate layers of convolutional neural network (CNN) and integrate them to derive the quality-aware feature representation. The quality evaluation module is composed of the classification sub-network for the true and pseudo 4K images and the regression sub-network for the quality assessment, and both of them adopt a multilayer perceptron (MLP) network to map the quality-aware features into the final quality index. The proposed model is trained through the multi-task learning manner. The uncertainty weighting method \cite{kendall2018multi} is adopted to balance the losses of these two tasks. The main contributions are as follows:

\begin{enumerate}
\item We propose an effective and efficient CNN based BIQA model to distinguish true and pseudo 4K images and assess the visual quality of 4K images, which is useful for UHD content producers and streaming service providers to evaluate and improve the QoE of UHD content.

\item We propose an effective texture-aware patch selection algorithm to select few local patches for the overall quality prediction of 4K content, which can reduce the computational overhead due to the high spatial resolution. The experimental results suggest that the local patches selected by our method are more representative to reflect the visual quality of the whole 4K image.

\item We propose a simple but effective CNN based feature extraction network which makes full use of visual information from low-level to high-level. What's more, the extracted quality-aware features are utilized to simultaneously recognize the true and pseudo 4K images and evaluate their visual quality via the multi-task learning manner.

\item Experimental results demonstrate that the proposed method outperforms all compared methods on four relevant 4K content quality assessment databases. The code of the proposed model will be publicly available.
\end{enumerate}

\section{Related work}
\label{sec2}

\subsection{BIQA for LR images}

Most existing BIQA methods are developed for LR images, whose resolution is normally no larger than 1080p. Here, we introduce some LR BIQA methods which are related to our study, including general-purpose BIQA methods, blur-specific BIQA methods, and SR BIQA methods.

\subsubsection{General-purpose BIQA}

General-purpose BIQA methods could be divided into three categories: natural scene statistics (NSS) based methods \cite{mittal2012no,fang2014no,liu2019unsupervised,gu2014hybrid,gu2016blind}, human visual system (HVS) based methods \cite{gu2014using,zhai2011psychovisual,zhai2019free,gu2013fisblim}, and learning-based methods \cite{sun2021blind,wang2021multi,lu2022cnn,gu2014deep,zhang2021uncertainty,zhang2020learning}. The NSS based methods are based on the principle that high quality natural scene images obey some statistical properties, while quality degradations lead to deviation from these statistics \cite{8648473}. Common NSS based methods include blind image spatial quality evaluator (BRISQUE) \cite{mittal2012no}, natural image quality evaluator (NIQE) \cite{mittal2012making}, integrated local NIQE (ILNIQE) \cite{zhang2015feature} and so on.  The mechanism of the HVS is an important source to design useful features for IQA. Zhai $et\ al$. \cite{zhai2011psychovisual} proposed a psychovisual quality metric in free-energy principle. Gu $et\ al$. \cite{gu2014using} modified the free-energy-based measure, and proposed an NR free energy-based robust metric (NFERM). Contrary to the conventional FR IQA metrics, Min $et\ al$. \cite{min2017blind,min2018blind} utilized a new “reference” called pseudo-reference image (PRI) and introduced a PRI-based BIQA framework. With the resurgence of machine learning research, many learning based BIQA methods were proposed in the last few years. A quality-aware clustering (QAC) method was introduced in \cite{xue2013learning}. Xu $et\ al$. \cite{xu2016blind} proposed a BIQA method based on high order statistics aggregation (HOSA).

\subsubsection{Blur-specific BIQA}

 Since blur is a common type of distortions widely encountered in image upscaling and super-resolution systems, we also introduce some BIQA methods for blur images. Narvekar and Karam \cite{5246972} employed a probabilistic model to estimate the probability of detecting blur at the image edges, then obtained a blur measure by pooling the cumulative probability of blur detection (CPBD). Vu and Chandler \cite{vu2012fast} developed a simple, yet effective wavelet-based algorithm for estimating both global and local image sharpness (FISH, Fast Image Sharpness). Li $et\ al$. \cite{li2015no} proposed a blind image blur evaluator (BIBLE) to assess image quality based on the variance-normalized moment energy.

\subsubsection{SR BIQA}

In recent years, numerous single-image SR (SISR) algorithms have been developed to HR images with their LR counterparts. At the same time, some SR IQA researches \cite{ma2017learning,beron2020blind,zhou2021image} have been proposed, which focus on the upsampling distortions introduced by the SISR algorithms. Ma $et\ al$. \cite{ma2017learning} proposed a BIQA metric for SR images, which is based on three types of NSS features in both spatial and frequency domains. Juan $et\ al$. \cite{beron2020blind} proposed an opinion-distortion unaware (ODU) BIQA model for SR IQA based on an optimal feature selection process, which picked out paired-product (PP) features and those derived from discrete cosine transform coefficients (DCT). These methods perform well on existing SR image datasets \cite{ma2017learning,zhou2019visual,kohler2019toward,beron2020blind} which only contain SR images up to 1080p, but are difficult to accurately predict the visual quality of real and fake 4K images, which is demonstrated in Section~\ref{4.2}.

\subsection{BIQA for 4K images}

Recently, two 4K BIQA works \cite{shah2021real,zhu2021perceptual} have been proposed. Rishi $et\ al$. \cite{shah2021real} developed a two-stage approach that classifies a video frame to have real or fake 4K resolution, where the first stage classifies local patches using a lightweight CNN, and the second stage aggregates local assessment into a global image level decision using logistical regression. Because the feature extraction module is relatively simple, Rishi's method lacks good generalizability to handle arbitrary upscaling methods. Moreover, Rishi's method cannot be used to evaluate the visual quality of 4K images. Zhu $et\ al$. \cite{zhu2021perceptual} proposed a BIQA metric to distinguish true and pseudo 4K content and measure the quality of 4K images. Specifically, frequency domain features and NSS based features of the images are extracted from three representative patches with high texture complexity, then the support vector regression (SVR) is employed to aggregate these features into the overall quality index. The method proposed by Zhu $et\ al$. performs well in the classification of true and pseudo 4K images, but the performance in quality regression of 4K images needs to be further improved. Both two methods achieve lower computational complexity via the use of local patches and are suitable for the real-time applications. However, there exist some limitations for these methods. First, the selected local patches are not representative enough to reflect the overall quality of 4K content, which is demonstrated in Section~\ref{4.3}. Then, the features extracted by a lightweight CNN or the hand-crafted features can not effectively represent the visual quality of 4K content.

\section{Proposed method}
\label{sec3}

\begin{figure*}[!htbp]
\centering
\includegraphics[width=6.5in]{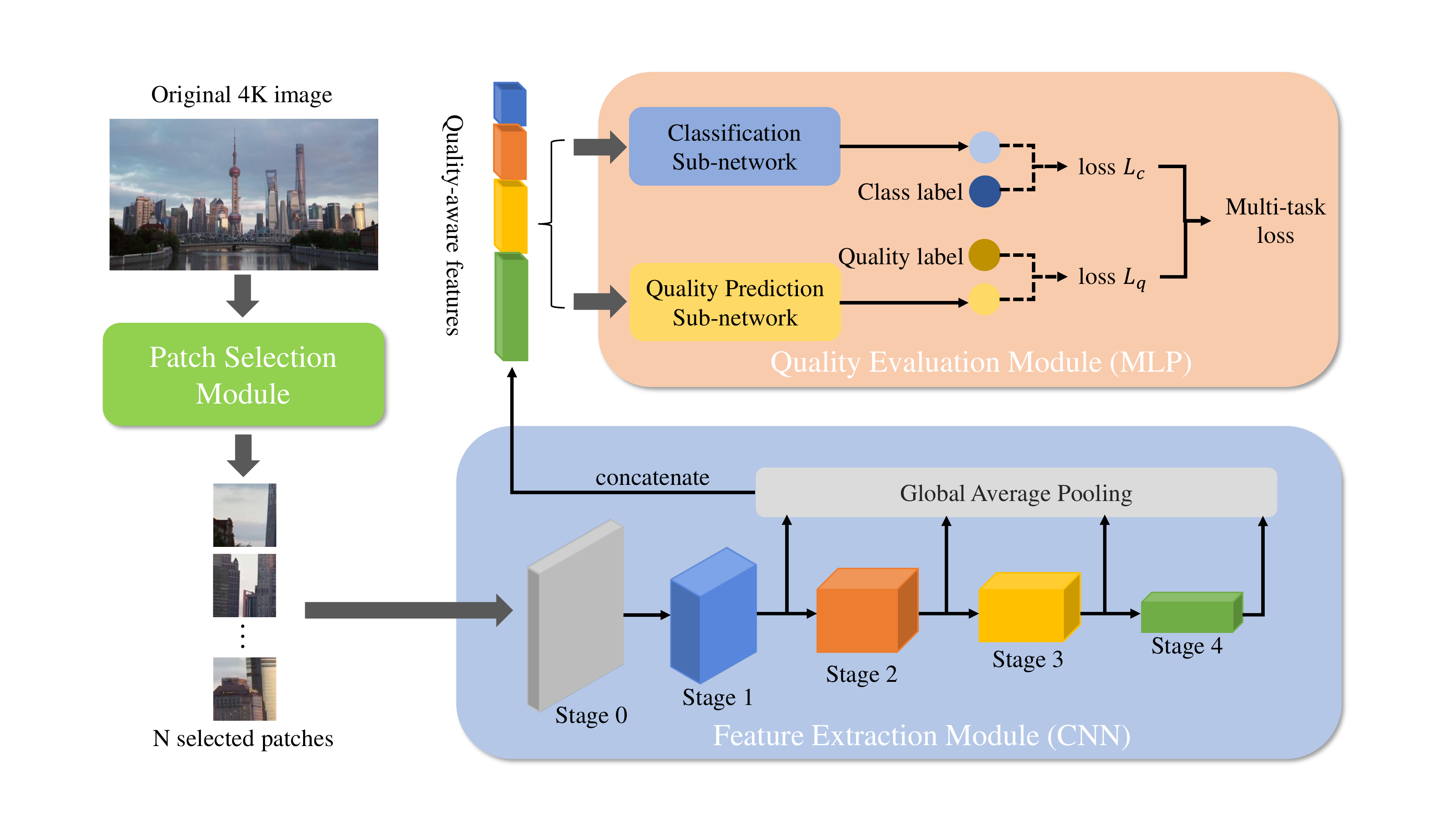}
\caption{The framework of the proposed BTURA model. The proposed model includes the patch selection module, the feature extraction module, and the quality evaluation module. The CNN is selected as the feature extraction backbone, which could be divided into five stages. The proposed model is built on a multi-task learning framework, in which the classification sub-network for true and pseudo 4K images and the image quality prediction sub-network share the feature extractor.}
\label{fig2.1}
\end{figure*}

In this section, we will introduce the proposed BTURA model in detail. The whole framework is illustrated in Fig.~\ref{fig2.1}. The proposed model can be divided into three modules, the patch selection module, the feature extraction module, and the quality evaluation module. First, we introduce how to select the representative patches from a 4K image. Then, we describe how to extract more effective quality-aware features for each patch through a CNN model. Finally, we will introduce how to obtain the class probabilities and quality scores of each patch and the whole 4K image through the multi-task learning manner.

\subsection{Patch selection module}
\label{patch_sel}

Due to the high computation complexity brought by large spatial resolution, it is not proper to extract features on the whole 4K image in practical applications. Also, high spatial resolution can represent more abundant high-frequency information, which is very important for texture-rich image regions to improve their visual quality. So, these texture-rich regions largely determine the visual quality of 4K content. 
Therefore, we propose a GLCM based texture complexity measure to select representative image patches from a 4K image.

Given a 4K image $I$, we first extract $M$ non-overlapped patches with a resolution of $S_{w} \times S_{h}$, where $M = \left \lfloor W/S_{w}\right \rfloor  \times \left \lfloor H/S_{h}  \right \rfloor$, $W$ and $H$ are the width and height of the 4K image, and $\left \lfloor \cdot  \right \rfloor $ refers to the round-down function. As the patches with high texture complexity can better reflect the characteristic of 4K images, the texture complexity of each patch is calculated to select representative patches, which could be expressed as:
\begin{equation}
t_{i} = {\rm F}(P_{i})    
\end{equation}
where ${\rm F}(\cdot)$ refers to texture complexity function and $t_{i}$ is the texture complexity value for patches $P_{i}, i = 1,2,3, \dots, M$. Finally, we choose the top $N$ patches $P_{1}, P_{2}, \dots, P_{N}$ with maximum texture complexity values:
\begin{equation}
P_{1}, P_{2}, \dots, P_{N} = \underset{P_{j_{1}}, P_{j_{2}}, \dots, P_{j_{N}}}{\operatorname{argmax}} {\rm F}(P_{j_{1}})+{\rm F}(P_{j_{2}})+\dots+{\rm F}(P_{j_{N}})
\end{equation}
where $P_{j_{1}}, P_{j_{2}}, \dots, P_{j_{N}}$ refer to $N$ different patches in the set $\{P_{i}\}$, $i \in 1,2,3,\dots,M$. As for the texture complexity function ${\rm F}(\cdot)$, we select the contrast of GLCM, which can effectively reflect the texture complexity \cite{mohanaiah2013image,7046402,barburiceanu20213d}.

GLCM describes the textures by measuring the spatial
correlation properties of gray scale. Given an image $I$ with a resolution of $W \times H$, the GLCM could be computed as:
\begin{equation}
\begin{gathered}
{\rm G L C M}(k, l)=\frac{1}{W \times H} \sum_{x} \sum_{y}\chi(x,y), \\
\chi(x,y)=\begin{cases}1, \text { if } I(x, y)=k \text { and } I(x+\Delta x, y+\Delta y)=l \\ 0,  \text { otherwise }\end{cases}
\end{gathered}
\end{equation}
where the pixel $(x+\Delta x, y+\Delta y)$ refers to the neighbor of the pixel $(x, y)$ in $I$ and their gray values are $I(x, y)$ and $I(x+\Delta x, y+\Delta y)$ ranging from 0 to 255. Then the texture complexity of an image is described by the contrast $f_{contrast}$ of the GLCM as follows:
\begin{equation}
    f_{contrast}=\sum_{k} \sum_{l}(k-l)^{2} {\rm G L C M}(k, l)
\end{equation}

\subsection{Feature extraction module}
\label{fem}

In recent years, CNN has demonstrated its powerful ability to solve various visual signal problems. Compared with the handcrafted features, the features extracted by CNN are more efficient and more suitable for various contents and distortions \cite{sun2021deep}. However, directly training a deep CNN model for BIQA methods may have some problems. Commonly adopted backbones such as VGG \cite{simonyan2014very} and ResNet \cite{he2016deep} are designed for image classification tasks and are used to extract high-level semantic features, but the perceived visual quality is also affected by low-level visual features. Therefore, it is not the best choice to directly use the popular CNN architecture as the backbone of the BIQA task.

Previous studies demonstrated \cite{zhao2018deepsim, ding2020image,zhang2018unreasonable,8603835} that the CNN can learn the meaningful visual features of the input data, and different layers of networks are able to learn different levels of features. The shallow convolution layers can learn the local characteristics of the image which reflect the low-level visual quality, while the deep convolution layers can learn more abstract features which represent the high-level semantic information. 
Inspired by this characteristic of CNN, we propose to utilize different kinds of features extracted from intermediate layers of the CNN model and integrate them into the quality-aware feature representation, which is depicted in Fig.~\ref{fig2.1}.

In the proposed feature extraction network, ResNet is adopted to extract features for image patches. Considering the computational efficiency and the predictive performance of the method, we choose ResNet-18 pre-trained on the ImageNet \cite{deng2009imagenet} as the backbone to extract features. The architecture of ResNet-18 can be divided into five stages, namely, Stage 0, Stage 1, Stage 2, Stage 3, and Stage 4. In each of the above stages except for Stage 0, there are several convolutional layers in series to deepen the network. The dimension of the feature maps at the current stage is half that of the previous stage while the number of channels is twice that of the previous stage. The feature maps Stage 1, Stage 2, Stage 3, and Stage 4 are extracted, referred to as $F_{1}$, $F_{2}$, $F_{3}$, and $F_{4}$, respectively. 
Then, these feature maps representing from low to
high level features are converted into feature vectors through the spatial global average pooling, which can be expressed as:
\begin{equation}
    f_{i}={\rm GAP}(F_{i}), i = 1, 2, 3, 4
\end{equation}
where ${\rm GAP}(\cdot)$ means the the spatial global average pooling and the dimension of converted feature vectors $f_{1}, f_{2}, f_{3}, f_{4}$ are 64, 128, 256 and 512, respectively.

Finally, the features of different levels are concatenated into the final quality-aware feature representation, which is expressed as:
\begin{equation}
    f_{q}=\operatorname{cat}\left(f_{1}, f_{2}, f_{3}, f_{4}\right)
\end{equation}
where $\operatorname{cat}(\cdot)$ denotes the concatenation operation and $f_{q}$ means the final 960-dimensional feature vector. 

\subsection{Quality evaluation module}

As mentioned in Section~\ref{patch_sel}, the selected $N$ representative image patches largely determine the visual quality of 4K content. Therefore, we only need to predict the quality indexes of the selected few patches, which can effectively reduce the computational complexity for 4K contents, and then obtain the quality index for the whole 4K content through the simple average pooling method. 

Specifically, with the extracted quality-aware features, two MLP networks are adopted to map them
into the class probabilities and quality score of each patch. The MLP network consists of two fully connected (FC) layers, where the first layer is composed of 128 neurons. For the classification sub-network ${\rm W}_{FC}^{C}(\cdot)$, the second layer contains 2 neurons to decide whether the images have true 4K resolution or not. For the quality evaluation sub-network ${\rm W}_{FC}^{R}(\cdot)$, the second layer contains 1 neuron to represent the visual quality of 4K images. Therefore, we can obtain the patch-level quality indexes via:
\begin{equation}
\begin{gathered}
    y={\rm W}_{FC}^{C}(f_{q}), \\
    q={\rm W}_{FC}^{R}(f_{q})
\end{gathered}
\end{equation}
where $y$ and $q$ denote the predicted positive probability and quality score of the selected patch. 
Then, the overall quality index of the 4K image is calculated as:
\begin{equation}
\begin{gathered}
    y_{pred} = \frac{1}{N}\sum_{n=1}^{N}y_{n}, \\
    q_{score} = \frac{1}{N}\sum_{n=1}^{N}q_{n}  
\end{gathered}
\end{equation}
where $y_{n}$ and $q_{n}$ are respectively the the predicted positive probability and quality score of the $n$-th selected patch $P_{j_{n}}$. $y_{pred}$ and $q_{score}$ are respectively the predicted positive probability and quality score  of the whole 4K image.

The proposed model is trained by the multi-task learning manner, which aims to improve prediction accuracy by learning the multiple regression and classification objectives from a shared representation. Specifically, two sub-tasks including the classification task of true and pseudo 4K images and the image quality prediction task are jointly optimized with a shared feature extraction module. The cross entropy is used as the loss function of the classification task, which can be computed as:
\begin{equation}
    L_{c} =-[y_{pred} \log y_{label} + (1-y_{pred}) \log (1-y_{label})]
\end{equation}
where $y_{label}$ is the ground-truth positive probability. As for the quality prediction task, we adopt the euclidean distance as the loss function:
\begin{equation}
    L_{q} =\left\|q_{score}-q_{mos}\right\|^{2}
\end{equation}
where $q_{mos}$ refers to the ground-truth quality score. For jointly optimizing the proposed BTURA model, an overall loss function should be defined. The relative weighting between each task’s loss has a significant impact on the performance of multi-task learning, while tuning these weights by hand is a difficult and expensive process \cite{kendall2018multi}. Hence, we adopt the task uncertainty weighting principle in \cite{kendall2018multi} to automatically weight multiple loss functions by considering the homoscedastic uncertainty \cite{kendall2017uncertainties} of each task. The multi-task loss function based on the uncertainty weighting principle is given as:
\begin{equation}
    L_{overall} = \frac{1}{2 \sigma_{1}^{2}} L_{c}+\frac{1}{2 \sigma_{2}^{2}} L_{q}+\log \sigma_{1}+\log \sigma_{2}
\end{equation}
where $\sigma_{1}$ and $\sigma_{2}$ are the learnable parameters related to the uncertainty of two sub-tasks. As the uncertainty decreases, the contribution of
the respective loss function increases. The last term in the equation acts as a regulariser for the parameter values. The overall loss is optimized with respect to model parameters as well as $\sigma_{1}, \sigma_{2}$, which can balance the losses $L_{c}$ and $L_{q}$ adaptively.

\begin{table*}[!ht]
\renewcommand\arraystretch{1.05}
\setlength\tabcolsep{2.5pt}
\normalsize
\centering
\caption{Summary of the test 4K databases.}\label{tab4.1}
\begin{tabular}{cccccccc}
\toprule%
Database & Contents & Scenes & Resolution & \makecell[c]{No. of upscaling\\methods} & \makecell[c]{Compression\\ methods} & Label & Range \\
\hline
4K IQA \cite{zhu2021perceptual} & 3,152 & 350 & 4K & 14 & - & MOS & [0,100]\\
BVI-SR \cite{mackin2018study} & 240 & 24 & 4K & 3 & - & MOS & [0,5]\\
MCML 4K UHD \cite{cheon2017subjective} & 250 & 10 & 4K & 1 & AVC, HEVC, VP9 & MOS & [0,10]\\
Waterloo IVC 4K \cite{li2019avc} & 1200 & 20 & 540P, 1080P, 4K & - & AVC, HEVC, VP9, AV1, AVS2 & MOS & [0,100]\\
\bottomrule
\end{tabular}
\end{table*}

\section{Experiment evaluation}
\label{sec4}

In this section, we first introduce the experiment protocol including the test databases, the evaluation metrics, the experimental setup, and the compared methods. Next, the comparison results between the proposed method and the compared state-of-the-art (SOTA) BIQA metrics on four 4K databases are reported. Then, we compare the performance of different texture complexity measures to verify the effectiveness of the GLCM based texture measure. Besides, we compare the computational complexity in terms of execution speed of all the methods and then conduct the ablation experiments to further validate the effectiveness of different kinds of features, multi-task learning, and task uncertainty weighting strategy. Finally, we analyze the effects of the number and size of patches on the performance of the proposed model.

\subsection{Experiment protocol}

\subsubsection{Test database}

The proposed method is mainly validated on four relevant databases: the 4K IQA database established by Zhu $et\ al$. \cite{zhu2021perceptual}, BVI-SR video quality database \cite{mackin2018study}, MCML 4K UHD video quality database \cite{cheon2017subjective}, and Waterloo IVC 4K video quality database \cite{li2019avc}, as summarized in Table~\ref{tab4.1}.

\begin{itemize}
    \item \textbf{4K IQA \cite{zhu2021perceptual}:} The 4K IQA database contains 3,152 4K images composed of 350 natural 4K images extracted from 4K video sequences and 2802 pseudo 4K images interpolated from 1080p and 720p images. The database not only has various contents but also is diverse and representative in upscaling operations, which involves 14 interpolation methods including traditional, deep learning-based, and video editing software and hardware integrated SR algorithms.
    \item \textbf{BVI-SR \cite{mackin2018study}:} The BVI-SR video quality database contains 240 five-second video sequences at 4K spatial resolution, including 24 source true 4K video sequences and 216 pseudo 4K video sequences with a range of spatial resolutions up to 4K. Three distinct spatial adaptation filters (including a CNN-based super-resolution method) are used for upsampling operations. 
    \item \textbf{MCML 4K UHD \cite{cheon2017subjective}:} The MCML 4K UHD video quality database consists of 10 pristine 4K sequences, which are encoded into 120 distorted 4K sequences and 120 distorted 1080P sequences by three video coding techniques: AVC \cite{wiegand2003overview}, HEVC \cite{sullivan2012overview}, and VP9 \cite{bankoski2013towards}. The compressed videos having 1080P resolution are up-scaled to the 4K resolution by the Lanczos resampling algorithm.
    \item \textbf{Waterloo IVC 4K \cite{li2019avc}:} The Waterloo IVC 4K video quality database is created from 20 pristine 4K videos and contains 1200 videos encoded by 5 encoders (AVC \cite{wiegand2003overview}, VP9 \cite{bankoski2013towards}, AV1 \cite{chen2018overview}, AVS2 \cite{gao2014overview}, and HEVC \cite{sullivan2012overview}) in 3 resolutions (540P, 1080P, 4K) and 4 distortion levels. We only test the performance of different methods on the 400 4K videos in the database.
\end{itemize}

It is noted that the true and pseudo 4K content refer to the original 4K content with high quality and the upscaled uncompressed 4K content respectively. Hence, as for the 4K IQA database and BVI-SR video quality database, which do not involves compression distortions, we validate the performance of the proposed method on both the classification and quality regression tasks. As for the other compressed 4K databases: MCML 4K UHD video quality database and Waterloo IVC 4K video quality database, we only preserve the quality prediction sub-network in the quality evaluation module of the proposed model and compare the performance of the proposed model and other methods in the quality regression task.


\begin{table*}[ht]
\renewcommand\arraystretch{1.05}
\normalsize
\caption{Performance comparison of the proposed model and thirteen BIQA methods on the 4K IQA database and BVI-SR video database, where $P_{T}$, $P_{F}$, $R_{T}$ ,$R_{F}$ refer to the precision and recall of positive and negative samples respectively. 
}\label{tab4.2}
\centering
\begin{tabular}{c|c|ccccccccc}
\toprule
\multirow{2}*{Type}& Dataset & \multicolumn{9}{c}{4K IQA}\\\cline{2-11}%
 & Criteria & SRCC  & KRCC  & PLCC   & RMSE    & $P_{T}$ & $P_{F}$ & $R_{T}$ & $R_{F}$ & Accuracy\\
\hline
\multirow{6}*{\makecell[c]{General-\\purpose}} & BRISQUE        & 0.6651 & 0.5061 & 0.6696 & 12.2003 & 0.8795       & 0.9477       & 0.5629    & 0.9904    & 0.9429 \\
~ & BMPRI          & 0.3594 & 0.2308 & 0.6534 & 12.4345 & 0.5890        & 0.9441       & 0.5486    & 0.9522    & 0.9074   \\
~ & HOSA           & 0.7153 & 0.5299 & 0.7173 & 11.4445 & 0.6613       & 0.9496       & 0.5914    & 0.9622    & 0.9210    \\
~ & NIQE           & 0.5223 & 0.3797 & 0.5691 & 13.5061 & 0.7550        & 0.9442       & 0.5371    & 0.9782    & 0.9293   \\
~ & NFERM          & 0.6708 & 0.4990  & 0.6662 & 12.2497 & 0.9653       & 0.9299       & 0.3971    & 0.9982    & 0.9315   \\
~ & QAC            & 0.6866 & 0.5204 & 0.6427 & 12.5836 & 0.4868       & 0.892        & 0.2114    & 0.9722    & 0.8877   \\
\hline
\multirow{3}*{\makecell[c]{Blur-\\specific}} & CPBD    & 0.5963 & 0.4315 & 0.6194 & 12.8950  & 0.5522       & 0.9349       & 0.4686    & 0.9525    & 0.8988   \\
~ & FISH  & 0.4775 & 0.3453 & 0.5781 & 11.1300   & 0.8764 & 0.8795 & 0.1414 & 0.9966 & 0.8792 \\
~ & BIBLE & 0.4419 & 0.3146 & 0.4692 & 11.8621 & 0.0000      & 0.8627 & 0.0000      & 1.0000      & 0.8627 \\
\hline
\multirow{2}*{SR} & Ma $et\ al$.  & 0.6497       & 0.4882     & 0.8481       & 7.0250        & 1.0000       & 0.9957       & 0.9729    & 1.0000    & 0.9963   \\
~ & Juan $et\ al$.    & 0.2561 & 0.1766 & 0.3020 & 12.7212  & 0.0000       & 0.8627       & 0.0000    & 1.0000    & 0.8627   \\
\hline
\multirow{3}*{4K} & TSARA  & -       & -       & -       & -        & 0.6757       & 0.9634       & 0.7086    & 0.9575    & 0.9299   \\
~ & Zhu $et\ al$.    & 0.8613 & 0.6927 & 0.8948 & 7.3045  & 0.9939       & 0.9908       & 0.9257    & 0.9993    & 0.9911   \\
~ & Proposed       & \bfseries0.9597 & \bfseries0.8509 & \bfseries0.9890  & \bfseries4.5537 & \bfseries1.0000       & \bfseries0.9984 & \bfseries0.9857 & \bfseries1.0000 & \bfseries0.9986\\
\midrule[1pt]
\multirow{2}*{Type}& Dataset & \multicolumn{9}{c}{BVI-SR}\\\cline{2-11}%
 & Criteria & SRCC  & KRCC  & PLCC   & RMSE    & $P_{T}$ & $P_{F}$ & $R_{T}$ & $R_{F}$ & Accuracy\\
\hline
\multirow{6}*{\makecell[c]{General-\\purpose}} & BRISQUE  & 0.8226 & 0.6418 & 0.8525 & 0.6690 & 0.0000 & 0.9000 & 0.0000 & \bfseries1.0000 & 0.9000 \\
~   & BMPRI    & 0.8249 & 0.6322 & 0.8718 & 0.6271 & \bfseries0.9000 & 0.9281 & 0.3000 & \bfseries1.0000 & 0.9300 \\
~   & HOSA     & 0.7478 & 0.5634 & 0.7493 & 0.8390 & 0.0000 & 0.9000 & 0.0000 & \bfseries1.0000 & 0.9000 \\
~   & NIQE     & 0.8185 & 0.6322 & 0.8476 & 0.6919 & 0.0000 & 0.9000 & 0.0000 & \bfseries1.0000 & 0.9000 \\
~   & NFERM    & 0.8195 & 0.6397 & 0.8502 & 0.6709 & 0.0000 & 0.9000 & 0.0000 & \bfseries1.0000 & 0.9000 \\
~   & QAC      & 0.3074 & 0.2217 & 0.3281 & 1.1975 & 0.0000 & 0.9000 & 0.0000 & \bfseries1.0000 & 0.9000 \\
\hline
\multirow{3}*{\makecell[c]{Blur-\\specific}}   & CPBD     & 0.4349 & 0.3258 & 0.3403 & 1.2387 & 0.0000 & 0.9000 & 0.0000 & \bfseries1.0000 & 0.9000 \\
~   & FISH     & 0.1454 & 0.1056 & 0.0987 & 1.3069 & 0.0000 & 0.9000 & 0.0000 & \bfseries1.0000 & 0.9000 \\
~   & BIBLE    & 0.0153 & 0.0016 & 0.5614 & 1.1336 & 0.0000 & 0.9000 & 0.0000 & \bfseries1.0000 & 0.9000 \\
\hline
\multirow{2}*{SR} & Ma       & 0.7748 & 0.5882 & 0.7718 & 0.8146 & 0.0000 & 0.9000 & 0.0000 & \bfseries1.0000 & 0.9000 \\
~  & Juan     & 0.3725 & 0.2809 & 0.3264 & 1.2327 & 0.0000 & 0.9000 & 0.0000 & \bfseries1.0000 & 0.9000 \\
\hline
\multirow{3}*{4K}  & TSARA    & -      & -      & -      & -     & 0.0000 & 0.9000 & 0.0000 & \bfseries1.0000 & 0.9000 \\
~   & Zhu      & 0.8266 & 0.6299 & 0.8843 & 0.5823 & 0.0000 & 0.9000 & 0.0000 & \bfseries1.0000 & 0.9000 \\
~  & Proposed & \bfseries0.8903 & \bfseries0.7295 & \bfseries0.9349 & \bfseries0.4226 & 0.7500 & \bfseries0.9565 & \bfseries0.6000 & 0.9778 & \bfseries0.9400\\
\bottomrule
\end{tabular}
\end{table*}

\subsubsection{Evaluation metrics}

In terms of the classification task for true and pseudo 4K images, Precision, Recall, and Accuracy are applied to evaluate the predictive performance of BIQA methods. Here positive and negative samples refer to true and pseudo 4K images, respectively. The values of these performance measures are between 0 and 1, the larger the values, the better the classification accuracy of the models. As for the regression task, Spearman Rank Order Correlation Coefficient (SRCC), Kendall's Rank Order Correlation Coefficient (KRCC), Pearson Linear Correlation Coefficient (PLCC), and Root Mean Squared Error (RMSE) are adopted to evaluate different BIQA methods. SRCC and KRCC represent the prediction monotonicity of the IQA method, PLCC reflects the prediction linearity and RMSE indicates the prediction accuracy. The value of
SRCC, KRCC, PLCC is between 0 and 1, and an excellent model is supposed
to obtain the values of these measures close to 1 and the value of RMSE close
to 0. Before calculating the PLCC and RMSE, we perform the nonlinear four-parameter logistic function in \cite{seshadrinathan2010study} to map the objective predictions to the subjective scores.

\subsubsection{Experiment setup}

In our experiments, as mentioned in Section~\ref{fem}, we use the ResNet-18 as the backbone of feature extraction module. We train and test the proposed model on a server with Intel Xeon Silver 4210R CPU @ 2.40 GHz, 128 GB RAM, and NVIDIA GTX 3090 GPU. The proposed model is implemented in PyTorch \cite{paszke2019pytorch}. The Adam optimizer \cite{kingma2014adam} with the initial learning rate 0.0002 is used to train the proposed model. The learning rate decays with a multiplicative factor of 0.9 for every 10 epochs and the epochs are set at 50. The batch size is set at 16. In the following experiments except for Section~\ref{ens}, we extract $16 \times 9$ non-overlapped patches with a resolution of $240 \times 240$ and then select three patches with high texture complexity, which means $S_{w}, S_{h}, M$, and $N$ are respectively set as 240, 240, 144 and 3. 

All test databases are divided into a training set of 80\% 4K contents and a test set of 20\% 4K contents. In order to ensure complete separation of training and testing contents, the 4K contents belonging to the same scene are assigned to the same set. The databases are randomly split 10 times, and the average value of the above evaluation criteria is computed as the final result.

\begin{table*}[ht]
\renewcommand\arraystretch{1.05}
\normalsize
\caption{Performance comparison of the proposed model and twelve BIQA methods on the MCML 4K UHD video quality database and Waterloo IVC 4K video quality database. 
}\label{tab4.3}
\centering
\begin{tabular}{c|c|cccc|cccc}
\toprule
\multirow{2}*{Type}& Dataset & \multicolumn{4}{c|}{MCML 4K UHD} & \multicolumn{4}{c}{Waterloo IVC 4K}\\\cline{2-10}%
 & Criteria & SRCC  & KRCC  & PLCC   & RMSE    & SRCC  & KRCC  & PLCC   & RMSE\\
\hline
\multirow{6}*{\makecell[c]{General-\\purpose}} & BRISQUE        & 0.6336 & 0.4746 & 0.5984 & 2.4166 & 0.4270 & 0.3146 & 0.3957 & 21.1158 \\
~ & BMPRI          & 0.4780 & 0.3535 & 0.4679 & 2.4970 & 0.3166 & 0.2255 & 0.3076 & 21.8007 \\
~ & HOSA           & 0.6983 & 0.5404 & 0.6936 & 2.4531 & 0.2798 & 0.1965 & 0.2673 & 22.1523 \\
~ & NIQE           & 0.5181 & 0.3874 & 0.5555 & 2.7032 & 0.3479 & 0.2434 & 0.2458 & 22.5107 \\
~ & NFERM          & 0.5940 & 0.4337 & 0.5112 & 2.4073 & 0.3325 & 0.2434 & 0.3266 & 21.9405 \\
~ & QAC            & 0.5803 & 0.4349 & 0.5189 & 2.7270 & 0.0595 & 0.0438 & 0.0528 & 22.5701 \\
\hline
\multirow{3}*{\makecell[c]{Blur-\\specific}} & CPBD    & 0.4837 & 0.3705 & 0.4431 & 2.6221 & 0.2657 & 0.1918 & 0.2603 & 22.2817 \\
~ & FISH  & 0.4813 & 0.3613 & 0.4638 & 2.6855 & 0.4063 & 0.2931 & 0.3150 & 21.9787 \\
~ & BIBLE & 0.4929 & 0.3820 & 0.4609 & 2.5655 & 0.2772 & 0.1980 & 0.2702 & 22.5830 \\
\hline
\multirow{2}*{SR} & Ma $et\ al$.  & 0.5522 & 0.4130 & 0.4799 & 2.6827 & 0.3810 & 0.2672 & 0.3615 & 22.3665 \\
~ & Juan $et\ al$.    & 0.3994 & 0.2860 & 0.5335 & 2.3830 & 0.3653 & 0.2662 & 0.3560 & 21.3251 \\
\hline
\multirow{2}*{4K} & Zhu $et\ al$.    & 0.5940 & 0.4337 & 0.5112 & 2.4073 & 0.4059 & 0.2894 & 0.3430 & 21.2782 \\
~ & Proposed       & \bfseries0.9353 & \bfseries0.8084 & \bfseries0.9608 & \bfseries0.6983 & \bfseries0.7393 & \bfseries0.5723 & \bfseries0.7709 & \bfseries13.3305\\
\bottomrule
\end{tabular}
\end{table*}

\begin{figure*}[ht]
\centering
\subfloat[]{\includegraphics[width=1.5in]{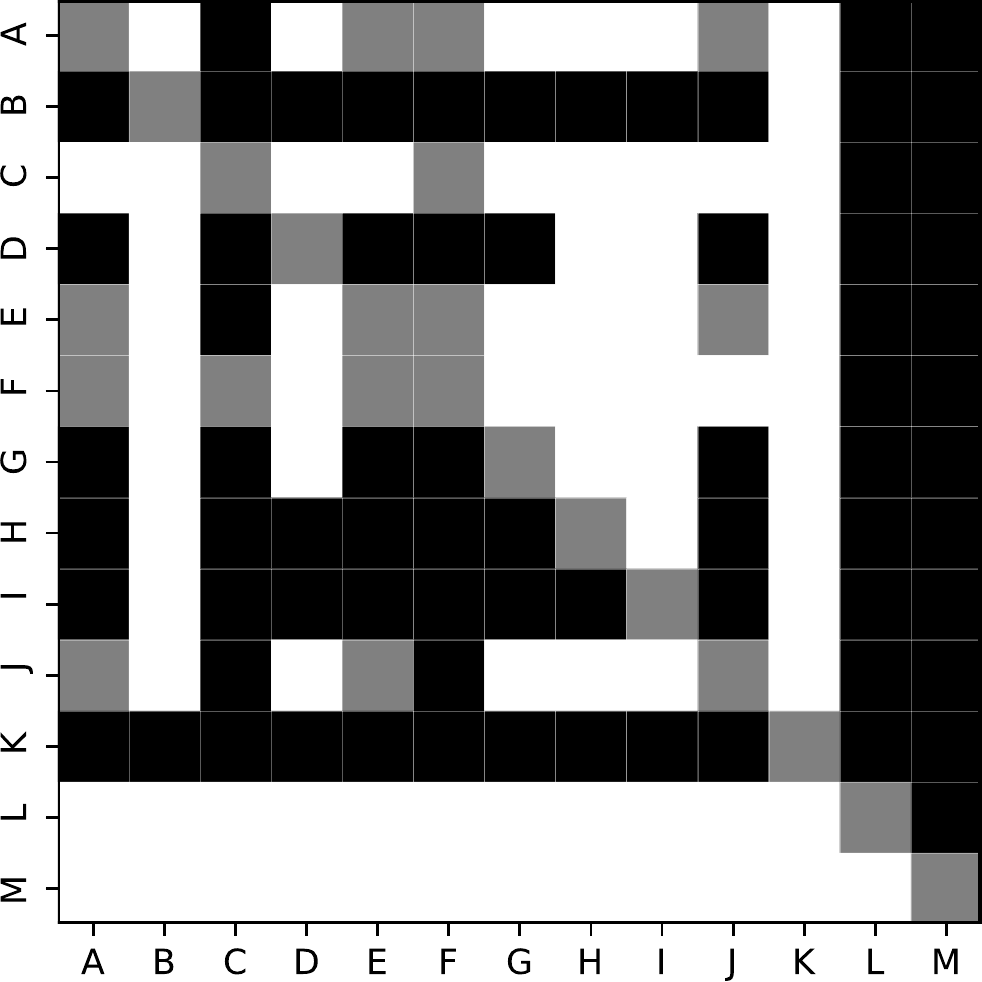}%
}
\hfil
\subfloat[]{\includegraphics[width=1.5in]{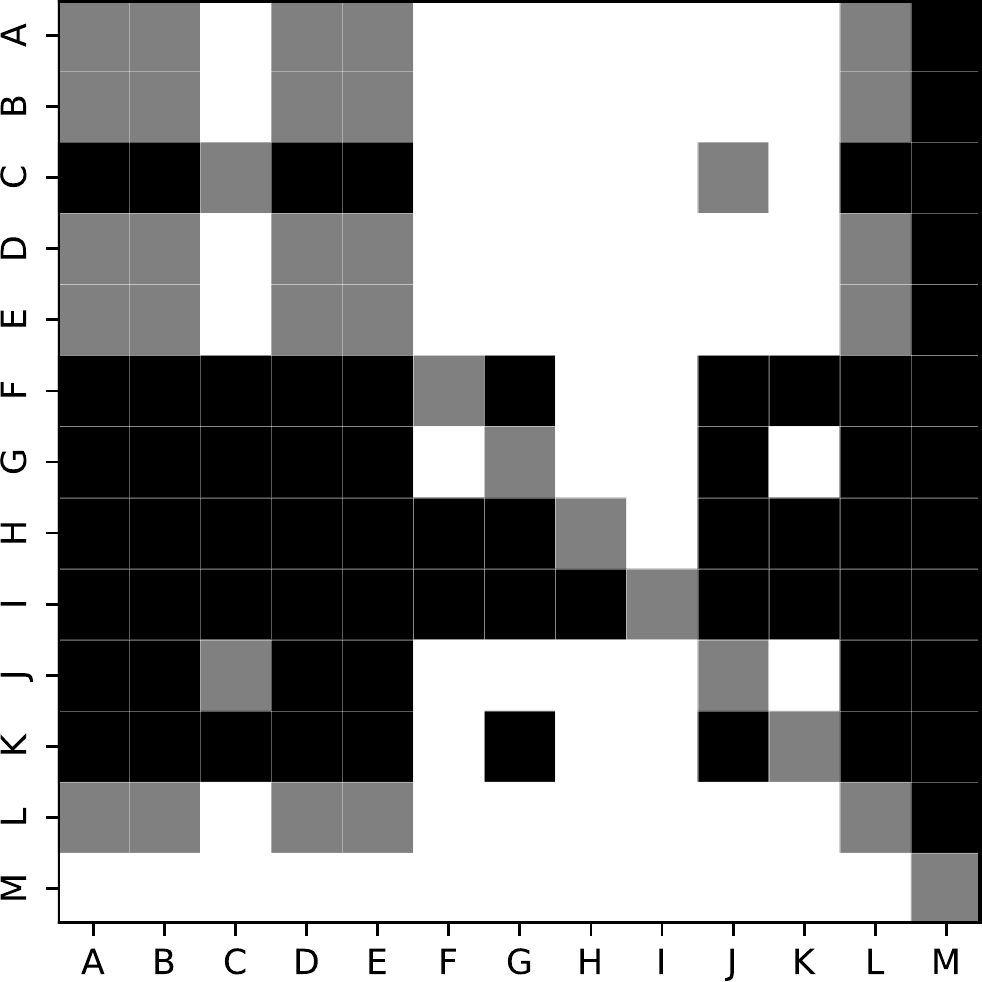}%
}
\hfil
\subfloat[]{\includegraphics[width=1.5in]{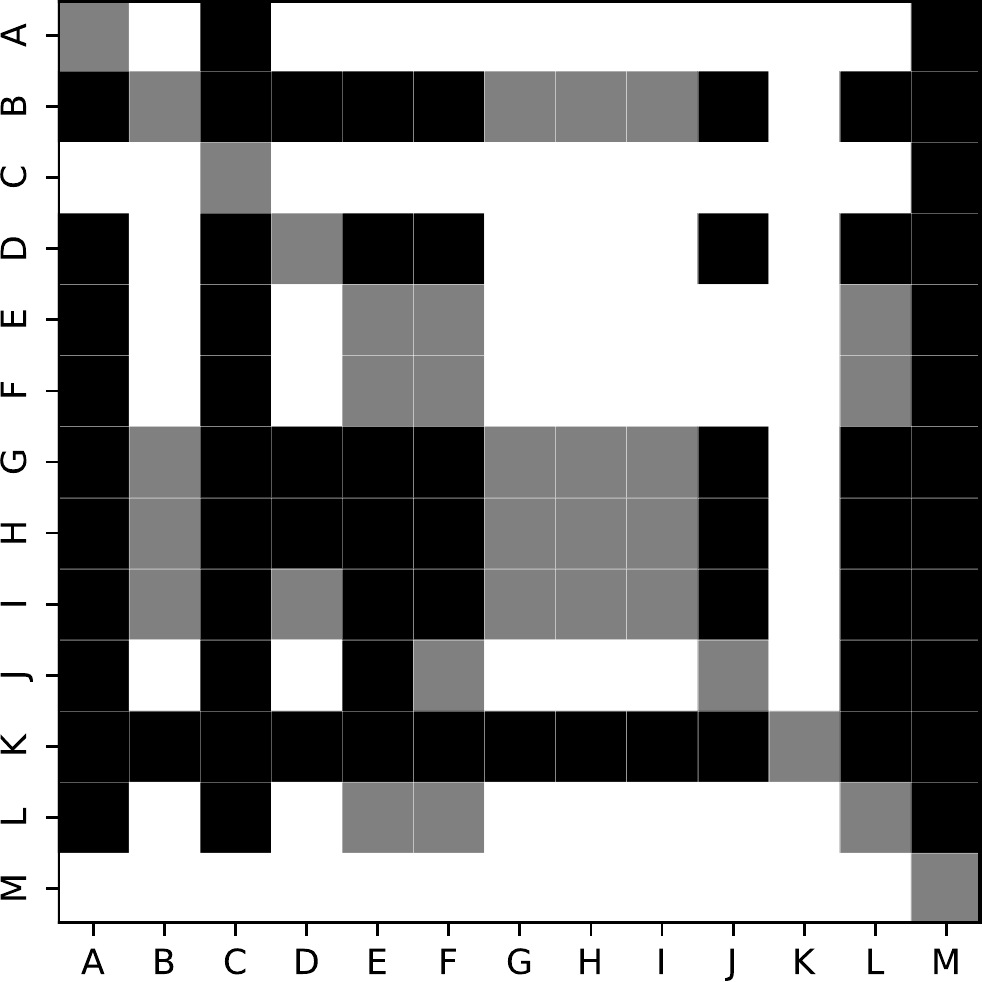}%
}
\hfil
\subfloat[]{\includegraphics[width=1.5in]{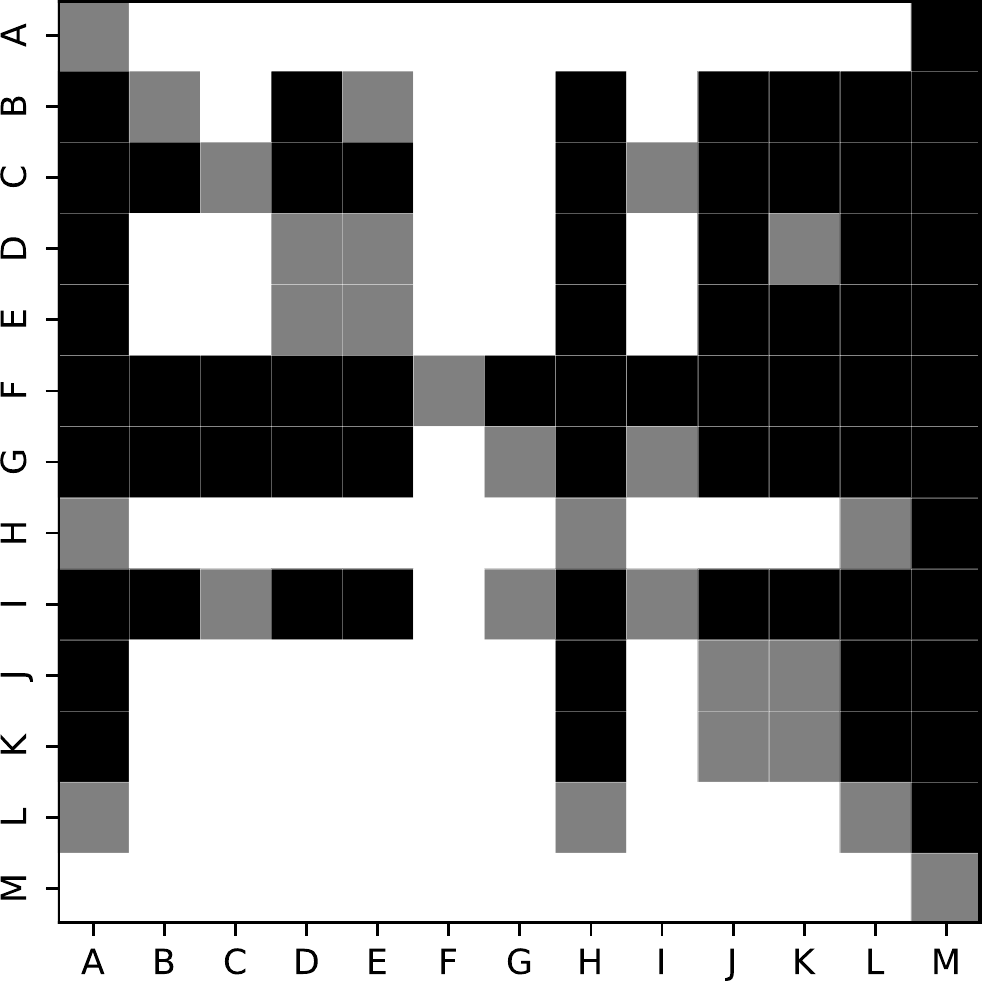}%
}
\caption{Statistical significance comparison between the proposed model and
other BIQA metrics on four test databases. A black/white block $(i,j)$ means
the method at row $i$ is statistically worse/better than the one at column $j$. A gray block $(m,n)$ means the method at row $m$ and the method at $n$ are statistically
indistinguishable. The metrics denoted by A-M are of the same order as the compared metrics except for TSARA in Table~\ref{tab4.2}.} \label{fig3.1}
\end{figure*}

\subsubsection{Compared methods}

We compare the proposed model with eleven popular BIQA methods for LR images and two SOTA 4K BIQA methods on four databases, including: 

\begin{itemize}
\item \textbf{General-purpose BIQA methods:} BRISQUE \cite{mittal2012no}, BMPRI \cite{min2018blind}, HOSA \cite{xu2016blind}, NIQE \cite{mittal2012making}, NFERM \cite{gu2014using}, and QAC \cite{xue2013learning}. 

\item \textbf{Blur-specific BIQA methods:} CPBD \cite{5246972}, FISH \cite{vu2012fast}, and BIBLE \cite{li2015no}. 

\item \textbf{SR BIQA methods:} The SR BIQA metric proposed by Ma $et\ al$. \cite{ma2017learning} and the SR BIQA model proposed by Juan $et\ al$. \cite{beron2020blind}. 

\item \textbf{4K BIQA methods:} The IQA metric proposed by Zhu $et\ al$. \cite{zhu2021perceptual} and TSARA \cite{shah2021real}. TSARA can only complete the classification task, so we could only compare its classification accuracy indicators with others. 
\end{itemize}

Among them, NIQE, CPBD, FISH, BIBLE, and the the SR BIQA model proposed by Juan $et\ al$. are opinion-unaware BIQA methods and do not need to be trained on the training set, while the remaining methods are retrained on the database. As for the three test 4K video databases, all the BIQA methods extract frame-level features or predict the frame-level results every half second, which are temporally average-pooled to obtain the overall video-level feature/result. 

\subsection{Performance comparison with other BIQA methods}
\label{4.2}

The performance of the proposed BIQA model and other thirteen BIQA methods on the four 4K database are listed in Table~\ref{tab4.2} and Table~\ref{tab4.3} respectively. The top 1 performance results in each column are marked in bold. From Table~\ref{tab4.2} and Table~\ref{tab4.3}, we can make the following conclusions. First, it is seen that the proposed model achieves SOTA or near-SOTA performance among all the compared BIQA metrics on all the test databases, which indicates that the proposed model is effective at predicting the quality of 4K contents. Next, since the 4K IQA database may have more diverse distortions brought by 14 types of upsampling operations than the BVI-SR database, which only involves three kinds of upsampling methods, most compared BIQA methods do not work well on the former but perform well on the latter. Some BIQA metrics based on handcrafted features perform better than TSARA, which adopts a lightweight CNN to extract features. It can be concluded that the features extracted by shallow CNN are not sufficient to deal with the distortions of pseudo 4K content caused by various SR algorithms. Then, compared with the quality regression task, the classification task of real and pseudo 4K contents is relatively easier. Most compared BIQA metrics do well in distinguishing the authenticity of 4K images on the 4K IQA database. Due to the unbalanced positive and negative samples in the BVI-SR database, many BIQA metrics are difficult to identify positive samples. 

\begin{figure*}[!htbp]
\centering
\subfloat[MOS=94.16, pred=95.62]{\includegraphics[width=2.15in]{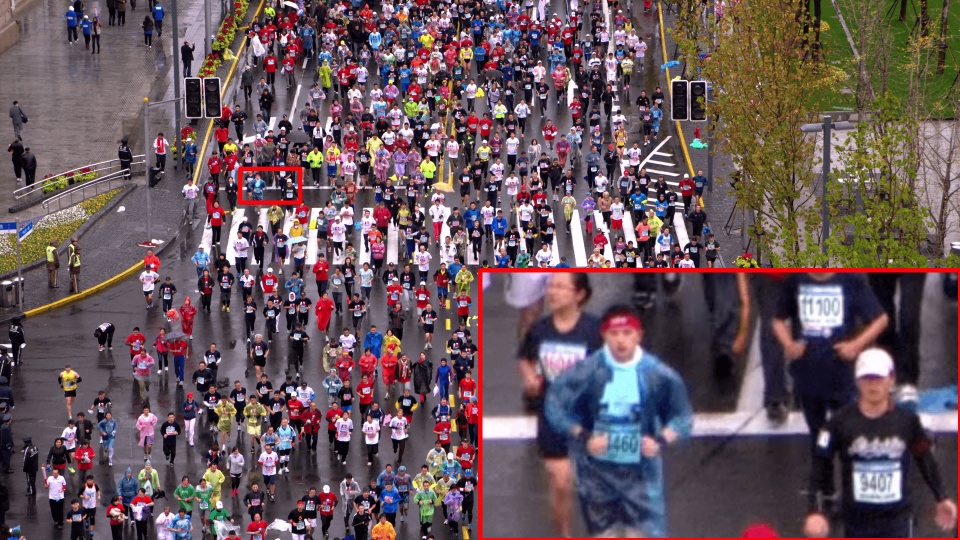}%
}
\hspace{3mm}
\subfloat[MOS=71.56, pred=69.75]{\includegraphics[width=2.15in]{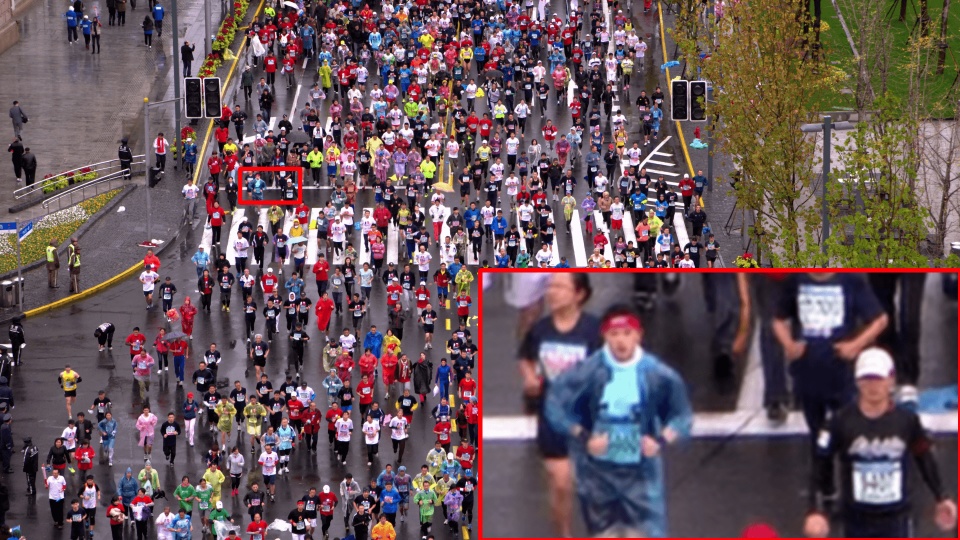}%
}
\hspace{3mm}
\subfloat[MOS=55.25, pred=55.16]{\includegraphics[width=2.15in]{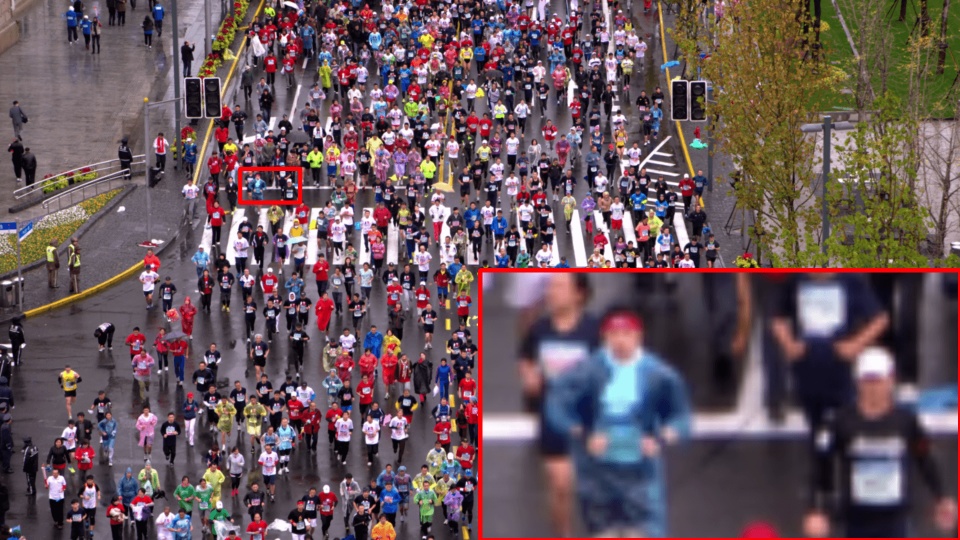}%
}
\quad
\subfloat[MOS=96.27, pred=97.80]{\includegraphics[width=2.15in]{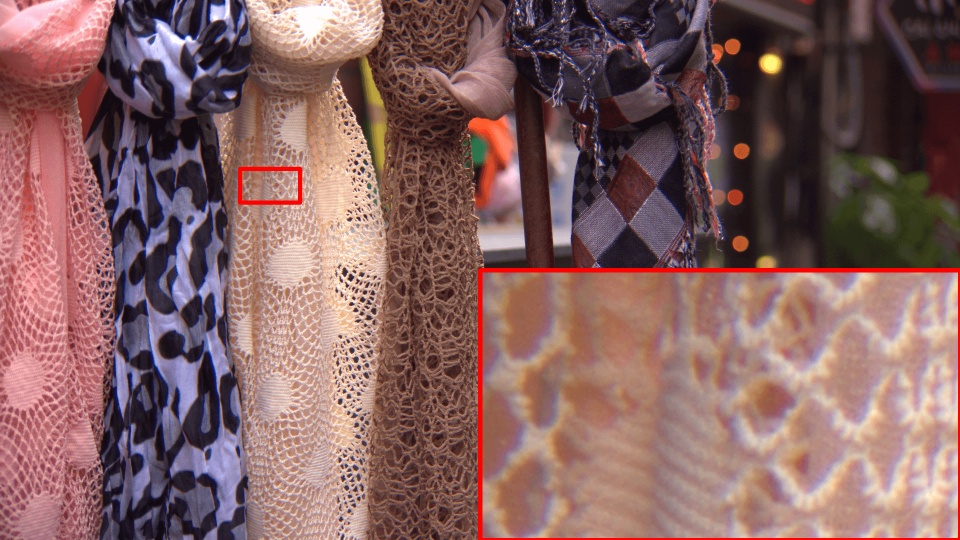}%
}
\hspace{3mm}
\subfloat[MOS=75.78, pred=79.02]{\includegraphics[width=2.15in]{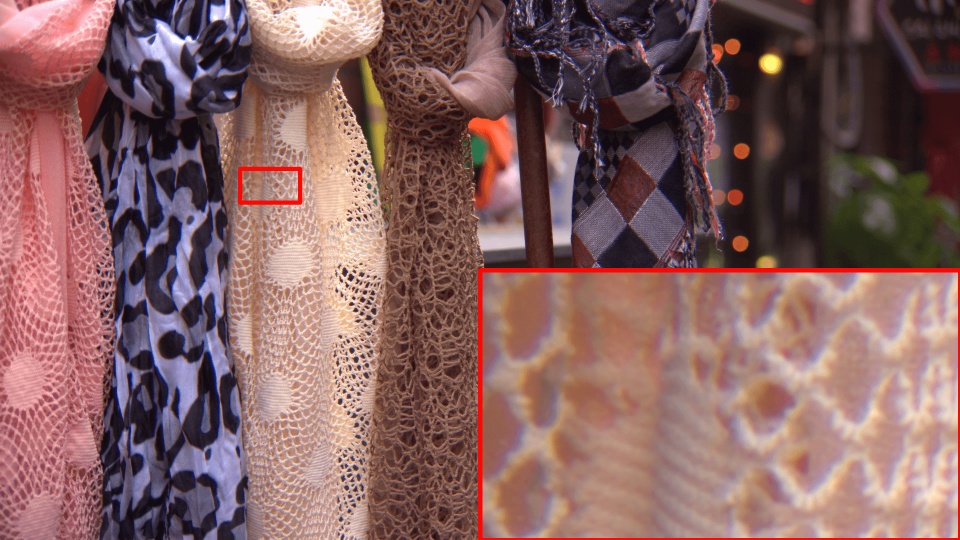}%
}
\hspace{3mm}
\subfloat[MOS=61.23, pred=60.83]{\includegraphics[width=2.15in]{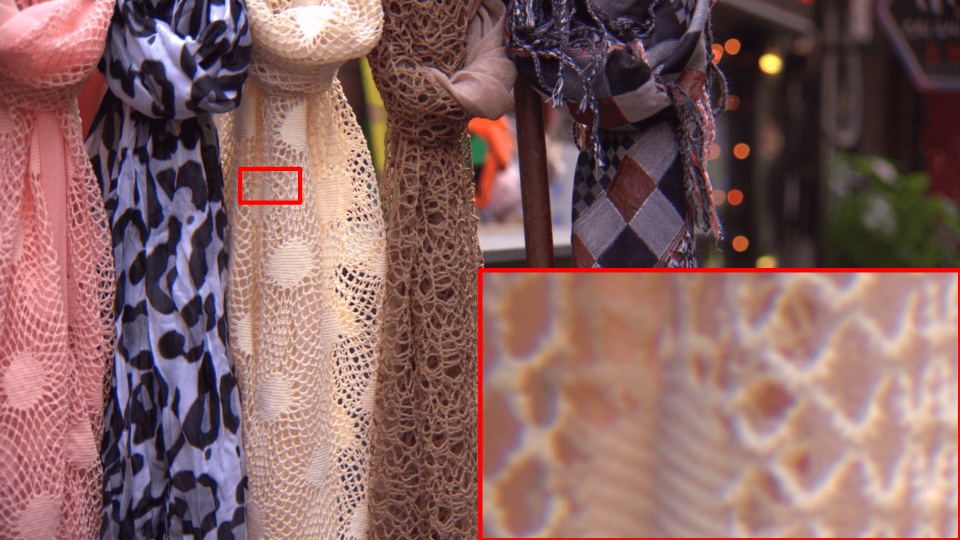}%
}
\quad
\subfloat[MOS=3.92, pred=3.81]{\includegraphics[width=2.15in]{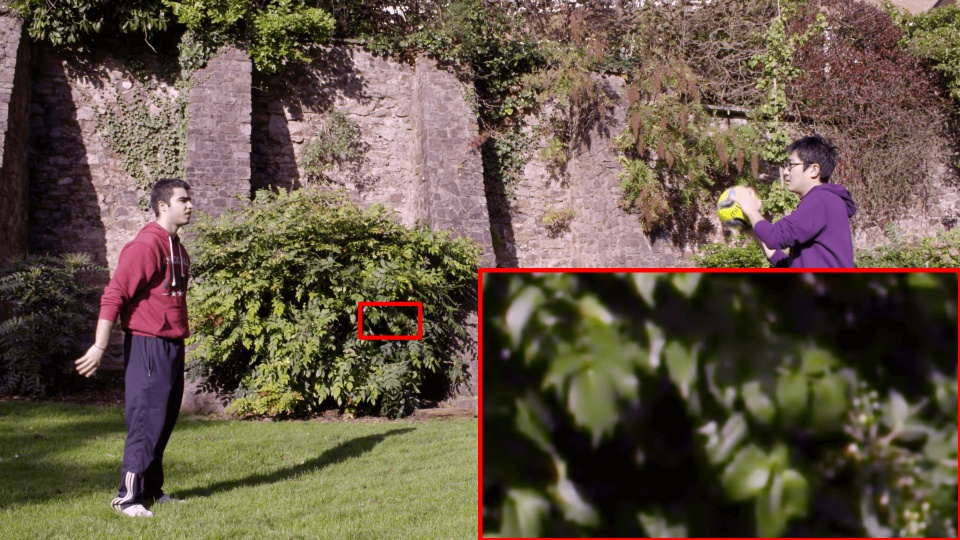}%
}
\hspace{3mm}
\subfloat[MOS=2.96, pred=3.03]{\includegraphics[width=2.15in]{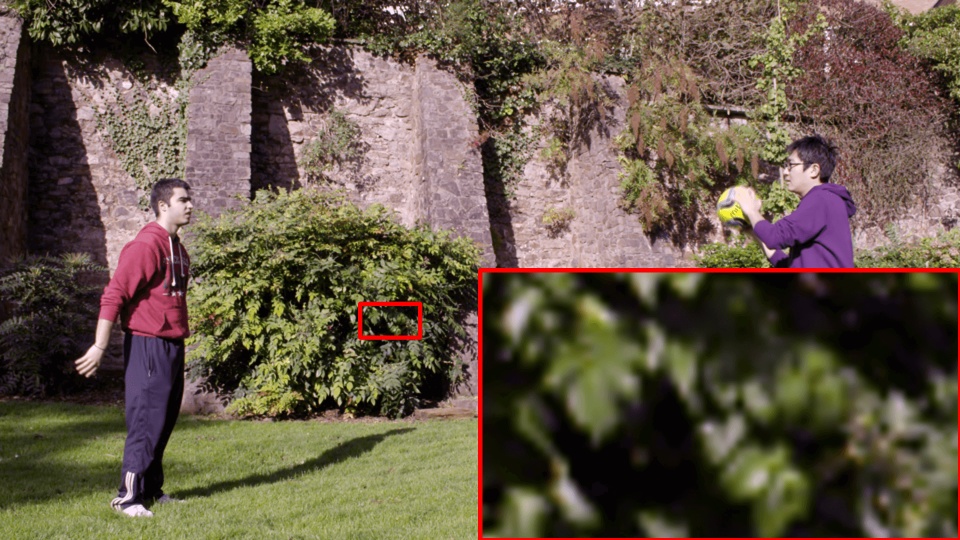}%
}
\hspace{3mm}
\subfloat[MOS=1.67, pred=1.58]{\includegraphics[width=2.15in]{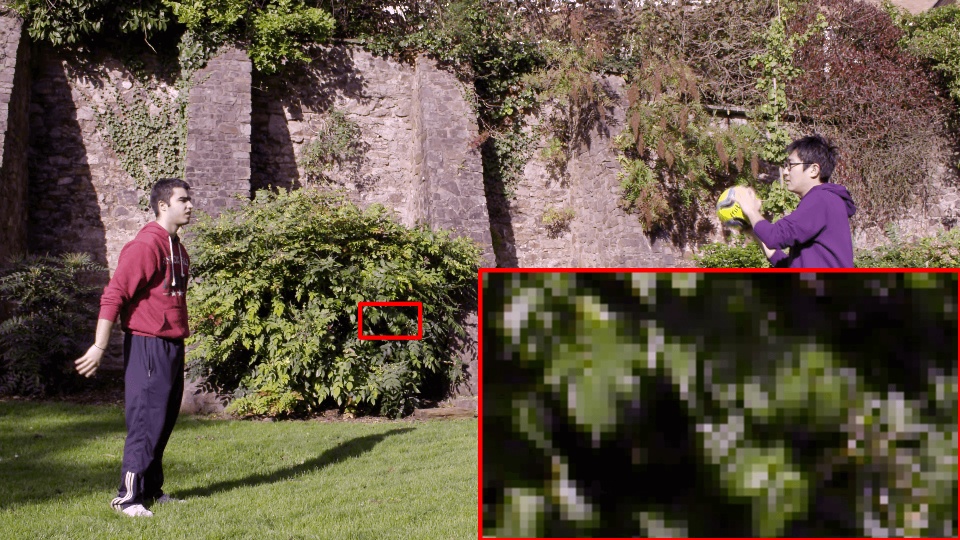}%
}
\quad
\subfloat[MOS=4.45, pred=4.33]{\includegraphics[width=2.15in]{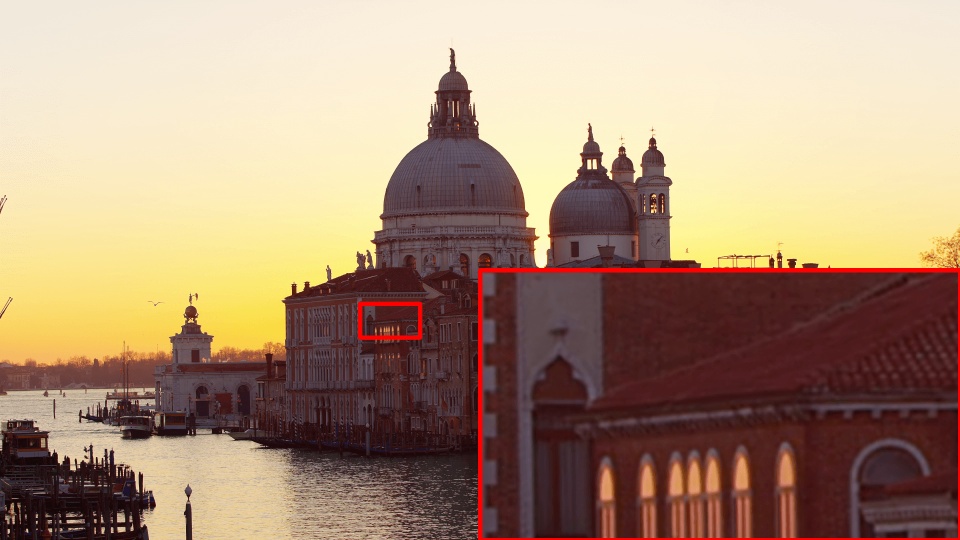}%
}
\hspace{3mm}
\subfloat[MOS=3.02, pred=2.96]{\includegraphics[width=2.15in]{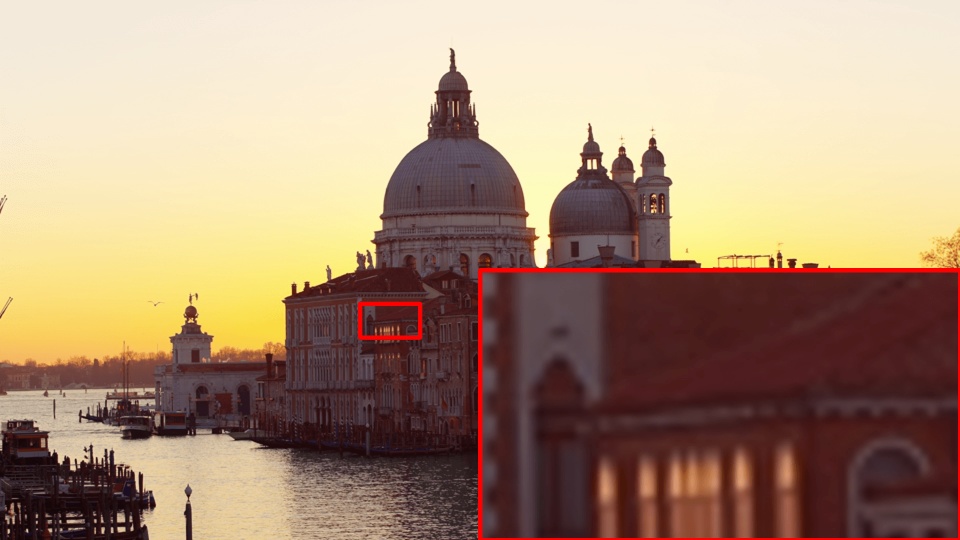}%
}
\hspace{3mm}
\subfloat[MOS=1.31, pred=1.52]{\includegraphics[width=2.15in]{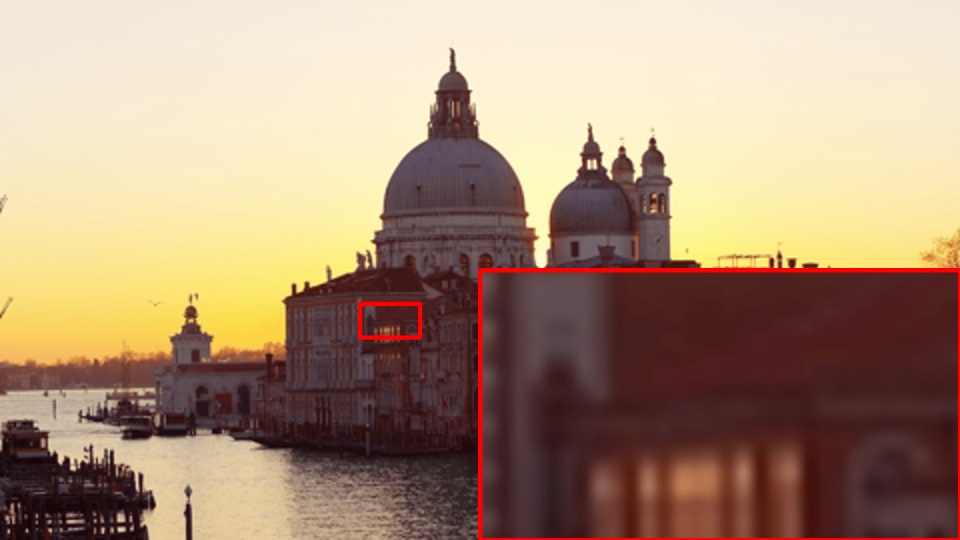}%
}
\caption{The predictive results of the proposed method for the example images or video frames. (a)-(f)  are from the 4K IQA database and (g)-(l) are from the BVI-SR database. The MOS (mean opinion score) and the predicted score are shown under each images, and the red rectangle region is cropped and enlarged for visualization. Notably, the MOSs in the 4K IQA database ranges from 0 to 100 while the MOSs in the BVI-SR database ranges from 0 to 5.}
\label{fig4}
\end{figure*}

Besides, two compared SR BIQA methods cannot accurately predict the perceptual quality of the true and pseudo 4K contents. The main reasons may be that these SR BIQA methods are designed for low-resolution SR images and the features extracted by them are not suitable for high-resolution images. Finally, it is worth mentioning that the proposed method achieves the best performance on both two compressed 4K video quality databases and leads by a large margin. This suggests that the proposed model has a strong ability to perceive the spatial distortios of 4K videos caused by various compression methods, despite facing the difficulty of quality prediction under unavailability of the reference signal.

To further analyze the performance of the proposed method and other BIQA metrics, we conduct the statistical significance test in \cite{sheikh2006statistical} to measure the difference between the predicted quality scores and the subjective ratings. Fig.~\ref{fig3.1} presents the results of the statistical significance test for the proposed method and other BIQA metrics on the four relevant 4K databases. We can clearly observe that the performance of our proposed 4K BIQA model is statistically superior to other compared BIQA metrics on all the test databases.

\begin{table*}[!ht]
\renewcommand\arraystretch{1.05}
\normalsize
\centering
\caption{Computational complexity of  different texture complexity measures.}\label{tab4.4}
\begin{tabular}{cccccc}
\toprule%
Measure & random & variance & local variance & entropy of hist & contrast of GLCM \\
\hline
Time(s) & - & 0.0878 & 0.3823 & \bfseries0.0697 & 0.1783\\
\bottomrule
\end{tabular}
\end{table*}

\begin{table*}[!ht]
\renewcommand\arraystretch{1.05}
\normalsize
\caption{Computational cost of the proposed model and other compared BIQA metrics.}\label{tab4.6}
\centering
\begin{tabular}{cccccccc}
\toprule%
Metric & BRISQUE & BMPRI & HOSA & NIQE & NFERM & QAC & CPBD \\
\hline
Time(s) & 1.6335 & 7.8568 & 0.6157 & 3.0467 & 650.1629 & 1.6380 & 11.0730\\
\hline
Metric & FISH & BIBLE & Ma $et\ al.$ & Juan $et\ al.$ & TSARA & Zhu $et\ al.$ & Proposed \\
\hline
Time(s) & 0.7243 & 11.7116 & 456.5131 & 215.5935 & \bfseries0.5482 & 0.6802 & 0.5749\\
\bottomrule
\end{tabular}
\end{table*}

\begin{figure}[!htbp]
\centering
\includegraphics[width=3.2in]{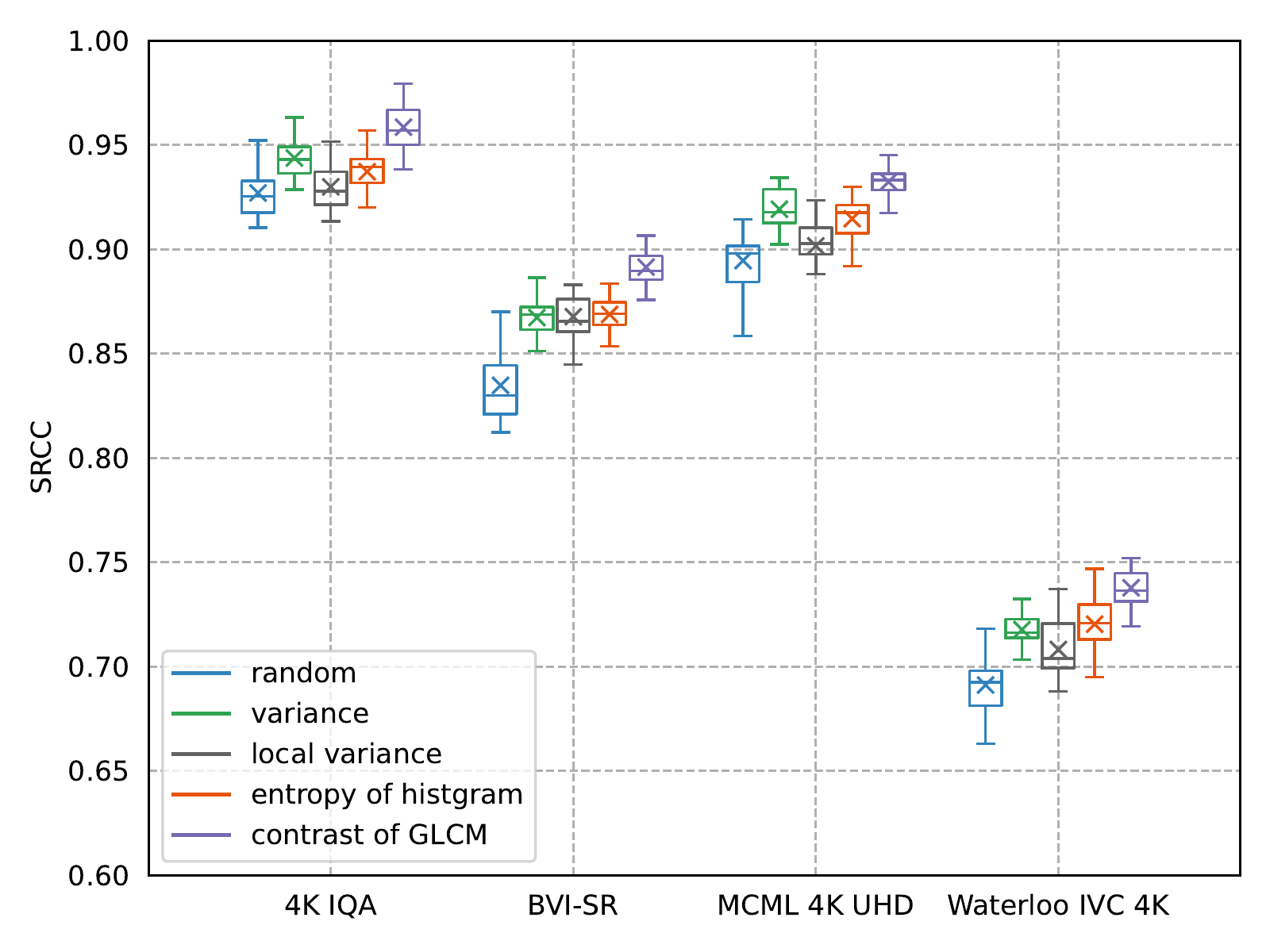}
\caption{Box plots for the performance comparison of texture complexity measures on all the test databases. Mean and standard deviation of SRCC values in 20 runs. The marks × in the middle represents
the average. The bottom, middle and top bounds of the box represent the 25\%,
50\% and 75\% percentage points, respectively}
\label{fig5}
\end{figure}

In addition, we select some 4K images or video frames in the 4K IQA database and BVI-SR database as examples to demonstrate the predictive performance of the proposed method. As shown in Fig.~\ref{fig4}, the difference between the subjective ground-truth scores (MOS) and the scores predicted by the proposed model is relatively small, which indicates that the proposed model performs well in the visual quality prediction of 4K contents.

\subsection{Performance of different texture complexity measures}
\label{4.3}

In this section, we mainly analyze the performance of different texture complexity measures in the patch selection module. In order to verify their effectiveness, the random selection method is used as the baseline. We select four effective and efficient texture measures, two of which are used in the previous 4K IQA work: variance \cite{shah2021real} and local variance \cite{zhu2021perceptual}, and the other two of which are entropy of gray difference histogram \cite{huang2010texture} and the proposed method: contrast of GLCM.
On the one hand, equipped with different texture measures, the final predictive performance of the whole BIQA model are compared, while the feature extraction module is set as the same as in Section~\ref{fem}. Here we only compare the method's performance in the quality prediction task. On the other hand, we compare the computational complexity of all the measures in terms of the execution speed, and the execution time is computed as the average time of 10 times measuring the texture complexity for 144 patches of a 4K image. We conduct the test on a server with Intel Xeon Silver 4210R CPU @ 2.40 GHz, 128 GB RAM.

The performance and computational complexity for different texture complexity measures are respectively depicted in Fig.~\ref{fig5} and Table~\ref{tab4.4}.  From Fig.~\ref{fig5}, we first can see that all the texture complexity measures perform better than the random method, which complies with the principle that the patches with rich textures are more important for the visual quality prediction of 4K contents. In addition, the contrast of the GLCM performs the best among these measures, which indicates that it is effective in describing texture complexity. Therefore, the contrast of the GLCM is employed in the patch selection module of our proposed method. From Table~\ref{tab4.4}, it is observed that the entropy of gray difference histogram outperforms compared methods in terms of computational cost and the contrast of the GLCM has a moderate computation cost. 

\subsection{Computational complexity of all the methods}

Considering the very high spatial resolution of 4K images, it is necessary to analyze the computational complexity  in terms of execution speed of BIQA methods, which is of great importance for real-time applications. In our experiment, since TSARA cannot predict the quality score of 4K images, the execution time is computed as the average time of 100 times classifying a 4K image to have real or fake 4K
resolution. We conduct the test on a Lenovo laptop computer with 1.8GHz
AMD Ryzen 7 4800U processor, 16 GB of RAM and Windows
10 operating system. For the experiments of compared BIQA methods except for TSARA, the software platform is MATLAB R2021a(9.10).

The computational cost of the proposed model and compared thirteen BIQA methods are listed in Table~\ref{tab4.6}. The top 1 performance results in each column are marked in bold.  From this table, we can observe that TSARA has the lowest computational cost because it classifies local patches through a lightweight CNN.  The three most time-consuming methods are respectively NFERM, Ma's method, and Juan's method, which evaluate the images through constructing a computationally expensive model. Both the BIQA method proposed by Zhu $et\ al.$ and the proposed method extract features on several selected patches and thus have relatively low computational cost. Besides, the proposed method outperforms all compared BIQA methods in terms of the predictive performance, which indicates that evaluating the visual quality of 4K images through several representative patches is an effective and efficient approach to reduce the computational cost.

\begin{table*}[ht]
\renewcommand\arraystretch{1.05}
\normalsize
\caption{Performance comparison of five models for ablation study, which is conducted to validated the effects of different kinds of features.}\label{tab4.8}
\centering
\begin{tabular}{c|c|cc|cc|cc|cc}
\toprule
\multirow{2}*{Backbone} & Dataset & \multicolumn{2}{c|}{4K IQA} & \multicolumn{2}{c|}{BVI-SR} & \multicolumn{2}{c|}{MCML 4K UHD} & \multicolumn{2}{c}{Waterloo IVC 4K}\\\cline{2-10}%
 & Criteria & SRCC  & PLCC   & SRCC & PLCC  & SRCC  & PLCC    & SRCC  & PLCC \\
\hline
\multirow{5}*{\makecell[c]{ResNet-18}} & BL  & 0.9490 & 0.9789 & 0.8784 & 0.9202 & 0.9261 & 0.9424  & 0.7239 & 0.7529 \\
~ & $\text{BL}_{234}$ & 0.9573 & 0.9874 & 0.8874 & 0.9330 & 0.9345 & 0.9575  & 0.7387 & 0.7688 \\
~ & $\text{BL}_{134}$ & 0.9558 & 0.9856 & 0.8866 & 0.9323 & 0.9312 & 0.9512  & 0.7369 & 0.7635 \\
~ & $\text{BL}_{124}$ & 0.9525 & 0.9765 & 0.8789 & 0.9237 & 0.9289 & 0.9455  & 0.7245 & 0.7554 \\
~ & $\text{BL}_{all}$ & \bfseries0.9597 & \bfseries0.9890 & \bfseries0.8903 & \bfseries0.9349 & \bfseries0.9353 & \bfseries0.9608  & \bfseries0.7393 & \bfseries0.7709 \\
\hline
\multirow{5}*{\makecell[c]{ResNet-50}} & BL  & 0.9472 & 0.9838 & 0.8827 & 0.9211 & 0.9285 & 0.9445  & 0.7214 & 0.7477 \\
~ & $\text{BL}_{234}$ & 0.9539 & 0.9862 & 0.8854 & \bfseries0.9234 & \bfseries0.9340 & \bfseries0.9597  & 0.7349 & 0.7612 \\
~ & $\text{BL}_{134}$ & 0.9532 & 0.9820 & 0.8866 & 0.9221 & 0.9325 & 0.9563 & 0.7333 & 0.7601 \\
~ & $\text{BL}_{124}$ & 0.9501 & 0.9847 & 0.8810 & 0.9199 & 0.9288 & 0.9451  & 0.7245 & 0.7511 \\
~ & $\text{BL}_{all}$ & \bfseries0.9580 & \bfseries0.9877 & \bfseries0.8870 & 0.9231 & 0.9331 & 0.9592  & \bfseries0.7354 & \bfseries0.7639  \\
\hline
\multirow{5}*{\makecell[c]{ResNeXt-50}} & BL  & 0.9499 & 0.9845 & 0.8810 & 0.9225 & 0.9294 & 0.9511 & 0.7229 & 0.7543 \\
~ & $\text{BL}_{234}$ & 0.9546 & \bfseries0.9886 & 0.8845 & 0.9266 & 0.9315 & 0.9567 & \bfseries0.7340 & 0.7639 \\
~ & $\text{BL}_{134}$ & 0.9485 & 0.9868 & 0.8832 & 0.9268 & 0.9303 & 0.9580 & 0.7312 & 0.7622 \\
~ & $\text{BL}_{124}$ & 0.9491 & 0.9876 & 0.8813 & 0.9230 & 0.9281 & 0.9497 & 0.7230 & 0.7568 \\
~ & $\text{BL}_{all}$ & \bfseries0.9591 & 0.9875 & \bfseries0.8848 & \bfseries0.9272 & \bfseries0.9323 & \bfseries0.9582 & 0.7331 & \bfseries0.7646 \\
\hline
\multirow{5}*{\makecell[c]{ResNet-101}} & BL  & 0.9398 & 0.9803 & 0.8698 & 0.9149 & 0.9134 & 0.9365 & 0.7156 & 0.7486 \\
~ & $\text{BL}_{234}$ & 0.9446 & 0.9835 & 0.8720 & \bfseries0.9201 & 0.9247 & 0.9481 & 0.7204 & \bfseries0.7555 \\
~ & $\text{BL}_{134}$ & 0.9464 & 0.9823 & 0.8714 & 0.9152 & 0.9240 & 0.9465 & 0.7203 & 0.7538 \\
~ & $\text{BL}_{124}$ & 0.9421 & 0.9817 & 0.8675 & 0.9160 & 0.9177 & 0.9389 & 0.7177 & 0.7489 \\
~ & $\text{BL}_{all}$ & \bfseries0.9485 & \bfseries0.9847 & \bfseries0.8731 & 0.9176 & \bfseries0.9258 & \bfseries0.9499 & \bfseries0.7211 & 0.7534  \\
\bottomrule
\end{tabular}
\end{table*}

\subsection{Ablation experiments}

In this section, in order to further validate the effectiveness of features extracted from different intermediate layers of CNN, the multi-task learning manner, and the task uncertainty weighting strategy, we conduct the following ablation experiments.

\subsubsection{Effects of different kinds of features}
Here we only compare the method's performance in the quality prediction task. We adopt ResNet-18, ResNet-50, ResNeXt-50 \cite{xie2017aggregated},and ResNet-101 as the backbone to construct our baseline model and analyze the performance of the proposed model without features from any intermediate layer. The five feature extraction models selected for the ablation experiment are referred to as:
\begin{itemize}
    \item BL: The baseline model.
    \item $\text{BL}_{234}$: The baseline model with features from Stage 2, Stage 3, and Stage 4
    \item $\text{BL}_{134}$: The baseline model with features from Stage 1, Stage 3, and Stage 4
    \item $\text{BL}_{124}$: The baseline model with features from Stage 1, Stage 2, and Stage 4
    \item $\text{BL}_{all}$: The proposed model.
\end{itemize}



The experimental results are listed in Table~\ref{tab4.8}. It is observed that the proposed model is superior to the other four models regardless of the employed backbone in most cases and the absence of features extracted from any one of the intermediate layers leads to performance degradation. This suggests that the features extracted from the intermediate layers of the CNN backbone can effectively reflect the visual quality of 4K contents. In terms of PLCC metric, the model $\text{BL}_{234}$ sometimes achieves the best prediction accuracy. However, the predictive monotonicity is more important than the prediction accuracy in practical applications. 

Then, among the five models, the baseline model performs worst on all the test databases, which means that it is not the best choice to directly use the popular CNN architecture for quality-aware feature extraction. Comparing the remaining models except for the baseline model, the experimental results indicate that the low-level features extracted from shallow convolution layers contribute relatively little to the high-level features extracted from deeper convolution layers. Finally, comparing the baseline models with different backbones (from ResNet-18 to ResNet-50, ResNeXt-50, and ResNet-101), we can see that the deeper network can not improve the performance in the quality prediction of 4K contents. Hence, the ResNet-18 is adopted as the backbone of the proposed method. 

\subsubsection{Effectiveness of the multi-task learning manner and the task uncertainty weighting strategy}

In this ablation experiment, the effectiveness of the multi-task learning manner and the task uncertainty weighting strategy in \cite{kendall2018multi} are further validated. Specifically, we compare the individual classification model and quality regression model to multi-task learning models using a unweighted sum of losses or the task uncertainty weighting strategy. The feature extraction module is set as the same as in Section~\ref{fem}.

\begin{table*}[!htbp]
\renewcommand\arraystretch{1.05}
\normalsize
\caption{Performance comparison of four models on the 4K IQA and BVI-SR database for ablation study, which is conducted to analyze the influence of multi-task learning and task uncertainty weighting strategy.}\label{tab4.9}
\centering
\begin{tabular}{c|cc|cc|cc}
\toprule%
\multirow{2}*{Model}& \multicolumn{2}{c|}{Task weights} & \multicolumn{2}{@{}c|@{}}{4K IQA} & \multicolumn{2}{c}{BVI-SR} \\\cline{2-7}%
  & Class.  & Regre.  &  SRCC & Accuracy  &  SRCC & Accuracy\\
\hline
Classification only               & 1 & 0 & - & \bfseries0.9996 & - & \bfseries0.9478  \\
Regression only & 0 & 1 & 0.9500 & - & 0.8840 & -  \\
Unweighted losses & 0.5 & 0.5 & 0.9545 & 0.9973 & 0.8868 & 0.9325 \\
Weighted losses & \checkmark   & \checkmark & \bfseries0.9597 & 0.9986 & \bfseries0.8903 & 0.9400   \\
\bottomrule
\end{tabular}
\end{table*}

From Table~\ref{tab4.9}, it is observed that the multi-task learning models with unweighted or weighted losses perform better than the individual quality regression model, which indicates that the quality prediction task could benefit from multi-task learning. Although the classification performance of the individual classification model is slightly better than that of the multi-task learning models, the image quality prediction task has broader application scenarios than the classification task. In addition, we can observe that the task uncertainty weighting strategy could improve the performance of the multi-task learning model, which means that correctly weighting loss terms is important for multi-task learning methods. 

\subsection{Effects of the number and size of patches}
\label{ens}

In this section, we mainly analyze how the number and size of selected patches with high texture complexity affect the final predictive performance of the proposed model. In terms of the number of patches, N is set as 3, 5, or 7. As for the width $S_{w}$ and height $S_{h}$ of selected patches, we consider three different sizes: $120 \times 120$, $240 \times 240$, and $480 \times 480$. The feature extraction module is set as the same as in Section~\ref{fem}. Here we only compare the method's performance in the quality regression task since it is more difficult.

\begin{table*}[!htbp]
\renewcommand\arraystretch{1.05}
\normalsize
\setlength\tabcolsep{5pt}
\caption{SRCC of the proposed model with different numbers and sizes of patches on all the test databases.}\label{tab4.7}
\centering
\begin{tabular}{c|ccc|ccc|ccc|ccc}
\toprule%
\multirow{2}*{$S_{w} \times S_{h}$} & \multicolumn{3}{c|}{4K IQA} & \multicolumn{3}{c|}{BVI-SR} & \multicolumn{3}{c|}{MCML 4K UHD} & \multicolumn{3}{c}{Waterloo IVC 4K} \\\cline{2-13}%
  & N=3  & N=5 & N=7   & N=3 & N=5  & N=7  & N=3 & N=5  & N=7 & N=3 & N=5  & N=7 \\
\hline
$120 \times 120$ & 0.9485 & 0.9504 & 0.9473 & 0.8815 & 0.8853 & 0.8845 & 0.9134 & 0.9180 & 0.9268 & 0.7199 & 0.7240 & 0.7266 \\
$240 \times 240$ & 0.9597 & 0.9611 & 0.9633 & 0.8903 & 0.8997 & 0.9117 & 0.9353 & 0.9380 & 0.9415 & 0.7393 & 0.7417 & 0.7479 \\
$480 \times 480$ & \bfseries0.9656 & 0.9627 & 0.9647 & 0.9012 & 0.9110 & \bfseries0.9124 & 0.9460 & 0.9498 & \bfseries0.9513 & 0.7505 & \bfseries0.7525 & 0.7520 \\
\bottomrule
\end{tabular}
\end{table*}

The results of SRCC on the four test databases are listed in Table~\ref{tab4.7}, and the top 1 performance results on each database are marked in bold. From Table~\ref{tab4.7}, we can see that the predictive performance is improved when the number of patches is fixed and the size of patches is increased. Specifically, the performance gain is most significant when the size of patches is increased from $120 \times 120$ to $240 \times 240$, which indicates that the size $120 \times 120$ is relatively small for the quality prediction task. Besides, it is observed that the performance improvement brought by the increases in the number of patches is not significant when the size of patches is fixed, which may indicate that three patches are relatively enough to accurately predicting the quality score of 4K contents. It is noted that the increases in the number and size of patches will lead to a high computational cost. Hence, to make a trade-off between the predictive performance and computational efficiency, the number and size of selected patches are respectively set as 3 and $240 \times 240$. 

\section{Conclusion}
\label{sec5}

With the growing popularity of 4K content, some low-quality pseudo 4K images and videos upscaled from LR deliver a very poor QoE to end-users. Hence, to monitor the perceptual quality of 4K content in multimedia industries, we propose an effective 4K BIQA model to recognize real and fake 4K images and evaluate the quality of 4K images. As the image regions with rich texture are important for evaluating the visual quality of HR images, we first select several representative patches from the whole 4K image by a GLCM based texture measure, which can greatly reduce the computation cost. Then a CNN based feature extraction network is adopted to extract quality-aware features for each patch. Specifically, the features extracted from different intermediate layers of CNN are concatenated into the final feature representation. Finally, two sub-networks are respectively utilized to map the quality-aware features into the patch-level class probabilities and quality scores, and the average pooling is adopted to obtain the overall quality index of the 4K image. The proposed model is trained through the multi-task learning manner and a task uncertainty weighting strategy is adopted to weight losses of two sub-tasks. The proposed model is validated on four 4K content databases, and the experimental results demonstrate that our proposed method outperforms the mainstream general-purpose and distortion-specific BIQA methods both in classification and regression tasks.

In summary, with the the patch selection strategy and the feature extraction module, the proposed model has a stronger ability to effectively predict the overall visual quality of 4K content compared with previous BIQA metrics. The proposed 4K BIQA method could not only help video service providers recognize original and pseudo 4K content, but also help to understand how different levels of space subsampling and compression affect the perceptual quality of videos and achieve better a trade-off between the perceptual quality of 4K content and transmission bandwidth.

\bibliographystyle{IEEEtran}
\bibliography{bibliography}

\vfill

\end{document}